\documentclass[acmtog]{acmart}
\acmSubmissionID{papers\_331s1}

\usepackage{booktabs} % For formal tables
\usepackage{url} % for website links
\usepackage{enumitem}
\usepackage{amsmath}
\usepackage{lipsum}

\acmPrice{15.00}

\setcopyright{acmlicensed}
\acmJournal{TOG}
\acmYear{2021}
\acmVolume{0}
\acmNumber{0}
\acmArticle{0}
\acmMonth{0}
\acmDOI{http://dx.doi.org/10.1145/8888888.7777777}

\setcitestyle{square}

\newcommand{\BW}[1]{{\color{orange}[Bruce: #1]}}

\newcommand{\PS}[1]{{\color{pink}[Pete: #1]}}

\newcommand{\REMOVED}[1]{}

\begin{document}

% Title. 
% If your title is long, consider \title[short title]{full title} - "short title" will be used for running heads.
\title{Efficient Dataflow Modeling of Peripheral Encoding in the Human Visual System}

\author{Rachel Brown}
\affiliation{%
  \institution{NVIDIA Research}}
\email{rabrown@nvidia.com}

\author{Vasha DuTell}
\affiliation{%
  \department{Vision Science}
  \institution{UC Berkeley}}
\email{vasha@berkeley.edu}

\author{Bruce Walter}
\affiliation{%
  \department{Computer Graphics}
  \institution{Cornell University}}
\email{bruce.walter@cornell.edu}

\author{Ruth Rosenholtz}
\affiliation{%
  \department{Brain and Cognitive Sciences, CSAIL}
  \institution{Massachusetts Institute of Technology}}
\email{rruth@mit.edu}

\author{Peter Shirley}
\affiliation{%
  \institution{NVIDIA Research}}
\email{pshirley@nvidia.com}

\author{Morgan McGuire}
\affiliation{%
  \institution{NVIDIA Research}}
\email{morgan3d@gmail.com}

\author{David Luebke}
\affiliation{%
  \institution{NVIDIA Research}}
\email{dluebke@nvidia.com}

% This command defines the author string for running heads.
% \renewcommand{\shortauthors}{DeJohnette, Rowland-Smith, Badeeri, and Foyt}
\renewcommand{\shortauthors}{Brown et al.}

% abstract
\begin{abstract}
Computer graphics seeks to deliver compelling images, generated within a computing budget, targeted at a specific display device, and ultimately viewed by an individual user.  The foveated nature of human vision offers an opportunity to efficiently allocate computation and compression to appropriate areas of the viewer's visual field, especially with the rise of high resolution and wide field-of-view display devices.  However, while the ongoing study of foveal vision is advanced, much less is known about how humans process imagery in the periphery of their vision--which comprises, at any given moment, the vast majority of the pixels in the image.  We advance computational models for peripheral vision aimed toward their eventual use in computer graphics.  In particular, we present a dataflow computational model of peripheral encoding that is more efficient than prior pooling-based methods and more compact than contrast sensitivity-based methods.  Further, we account for the explicit encoding of ``end stopped'' features in the image, which was missing from previous methods.  Finally, we evaluate our model in the context of perception of textures in the periphery.  Our improved peripheral encoding may simplify development and testing of more sophisticated, complete models in more robust and realistic settings relevant to computer graphics. 
\end{abstract}

%CCS
\begin{CCSXML}
<ccs2012>
   <concept>
       <concept_id>10010147.10010371.10010387.10010393</concept_id>
       <concept_desc>Computing methodologies~Perception</concept_desc>
       <concept_significance>500</concept_significance>
       </concept>
   <concept>
       <concept_id>10010147.10010371.10010382</concept_id>
       <concept_desc>Computing methodologies~Image manipulation</concept_desc>
       <concept_significance>300</concept_significance>
       </concept>
   <concept>
       <concept_id>10010147.10010371.10010395</concept_id>
       <concept_desc>Computing methodologies~Image compression</concept_desc>
       <concept_significance>300</concept_significance>
       </concept>
   <concept>
       <concept_id>10010147.10010341</concept_id>
       <concept_desc>Computing methodologies~Modeling and simulation</concept_desc>
       <concept_significance>100</concept_significance>
       </concept>
 </ccs2012>
\end{CCSXML}

\ccsdesc[500]{Computing methodologies~Perception}
\ccsdesc[300]{Computing methodologies~Image manipulation}
\ccsdesc[300]{Computing methodologies~Image compression}
\ccsdesc[100]{Computing methodologies~Modeling and simulation}

%keywords
\keywords{human vision, perception, foveated rendering, image compression}

% A "teaser" figure, centered below the title and authors and above the body of the work.
\begin{teaserfigure}
  \centering
  \includegraphics[width=1\textwidth]{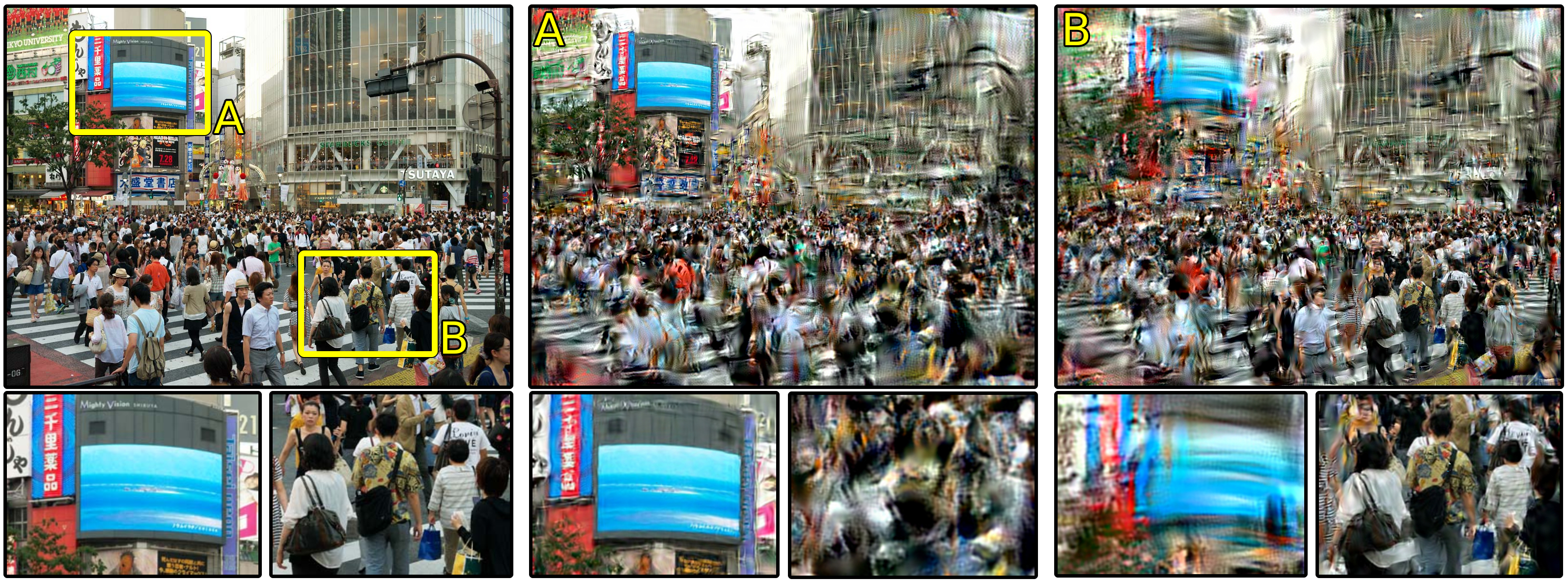}
  \caption{An example of our peripheral encoding model applied to a high-resolution color image for two different gaze locations (outlined in yellow) illustrates the effects of crowding. Despite their scrambled appearance, at an ideal viewing distance the middle and right images would be indistinguishable from the original image on the left, when focusing on the blue screen and the Hawaiian shirt, respectively. The bottom row shows zoomed-in cutouts of both gaze locations in each image, clearly demonstrating that peripheral degradation due to crowding is not merely the result of decreased resolution. We propose an efficient model for generating images that mimic the effect of crowding for use in a variety of graphics and vision science applications. Best viewed electronically at full-resolution.}
\end{teaserfigure}
\maketitle
\section{Introduction}

Computer graphics traditionally renders images with equal detail across the field of view, but researchers have long recognized that a viewer can only perceive a limited amount of this detail at a given moment.  The human visual system concentrates sensor resolution, encoding, and processing on the small region of the retina, called the \emph{fovea}, at the center of the viewer's gaze. The characteristics and limitations of foveal vision are very well studied, and the design of displays and rendering algorithms reflects this understanding. However, nearly all the pixels of a high-resolution image fall in the user's peripheral vision at a given moment; the foveal region subtends less than 5\% of a 15" laptop screen at typical distances, and less than 0.5\% of a modern virtual reality headset. This affords opportunities to reduce computation using \textit{foveated rendering}, (e.g., Kaplanyan et al.~\shortcite{kaplanyan2019deepfovea}, Patney et al.~\shortcite{patney2016towards}, Stengel and Magnor ~\shortcite{stengel2016gaze}, Guenter et al. \shortcite{guenter2012foveated}, Cater et al.~\shortcite{cater2002selective}, Luebke and Hallen~\shortcite{luebke2001perceptually}, Reddy~\shortcite{reddy2001perceptually}), or reduce display bandwidth or pixels using \emph{foveated display} (e.g., ~\cite{kim2019foveated}, ~\cite{vieri201818}, ~\cite{reder1973line}, ~\cite{basinger1982technical}). Such systems use both gaze tracking and models of peripheral vision to achieve lower rendering or resolution costs while also producing a percept that is identical to that of a full-detail image. As gaze tracking hardware becomes common, foveated techniques have the potential to reduce graphics computation by one or more orders of magnitude - savings that can be channeled into more realistic images, improved power consumption, cost, or form factor for devices, or reduced bandwidth across display busses or wireless networks. 

However, foveated graphics techniques to date leave something on the table: they do not explicitly account for the nature of \emph{peripheral encoding}, an important source of potential savings. Graphics researchers have largely built on models of peripheral vision \textit{acuity} that account for the resolution of photoreceptors on the retina, such as the well-studied spatial contrast sensitivity function~\cite{rovamo1978cortical}.  But human peripheral vision is not simply a low-resolution version of foveal vision, and as we show below, acuity alone does not account for the limitations and opportunities inherent in foveated rendering. 

In recent years, vision science researchers have made great strides toward understanding peripheral encoding, accounting for the dominant effect in peripheral vision known as \textit{crowding}, the visual system's vulnerability to clutter. However, this understanding has not been operationalized into efficient models that can guide graphics and display tasks such as rendering or compression. Our work advances state of the art in peripheral encoding, making it faster and easier to develop more sophisticated models that explain a wider range of visual stimuli, and to test those models in a setting that is more robust, realistic, and relevant to the graphics community. By creating more effective tools for vision science, we hope to pave the way for vision research to better inform graphics pipelines that take full advantage of the limitations of human vision.

To this end, we present a novel and efficient dataflow model that computes the encoding of peripheral imagery and captures recent vision science understanding of peripheral encoding, including accounting for ``end-stopped'' image features~\cite{zetzsche1990fundamental}.  Though not a neural network, our model uses the high-performance optimization machinery of modern deep learning and runs orders of magnitude faster than previous work.  We evaluate the model with a psychophysical study on perception of textures in the visual periphery. We believe our model will enable new research on perceptually-adaptive graphics, and hope it also serves as a call to action for further research and refinement of perceptual encoding models. 

Our contributions:
\begin{itemize}[topsep=0pt]
    \item a survey of current knowledge about peripheral perceptual encoding (Sec.~\ref{sec:background}), 
    \item a novel perceptual encoding model that flexibly allows inclusion of new image features such as end stopping, and is geared toward an efficient dataflow implementation (Sec.~\ref{sec:model}),
    \item a new data set of rendered textures and modernized texture descriptors (Sec.~\ref{sec:data}), and
    \item a psychophysical experiment on texture perception that illustrates the encoding model in action (Sec.~\ref{sec:evaluation}).
\end{itemize}

\begin{figure}
\includegraphics[width=0.99\columnwidth]{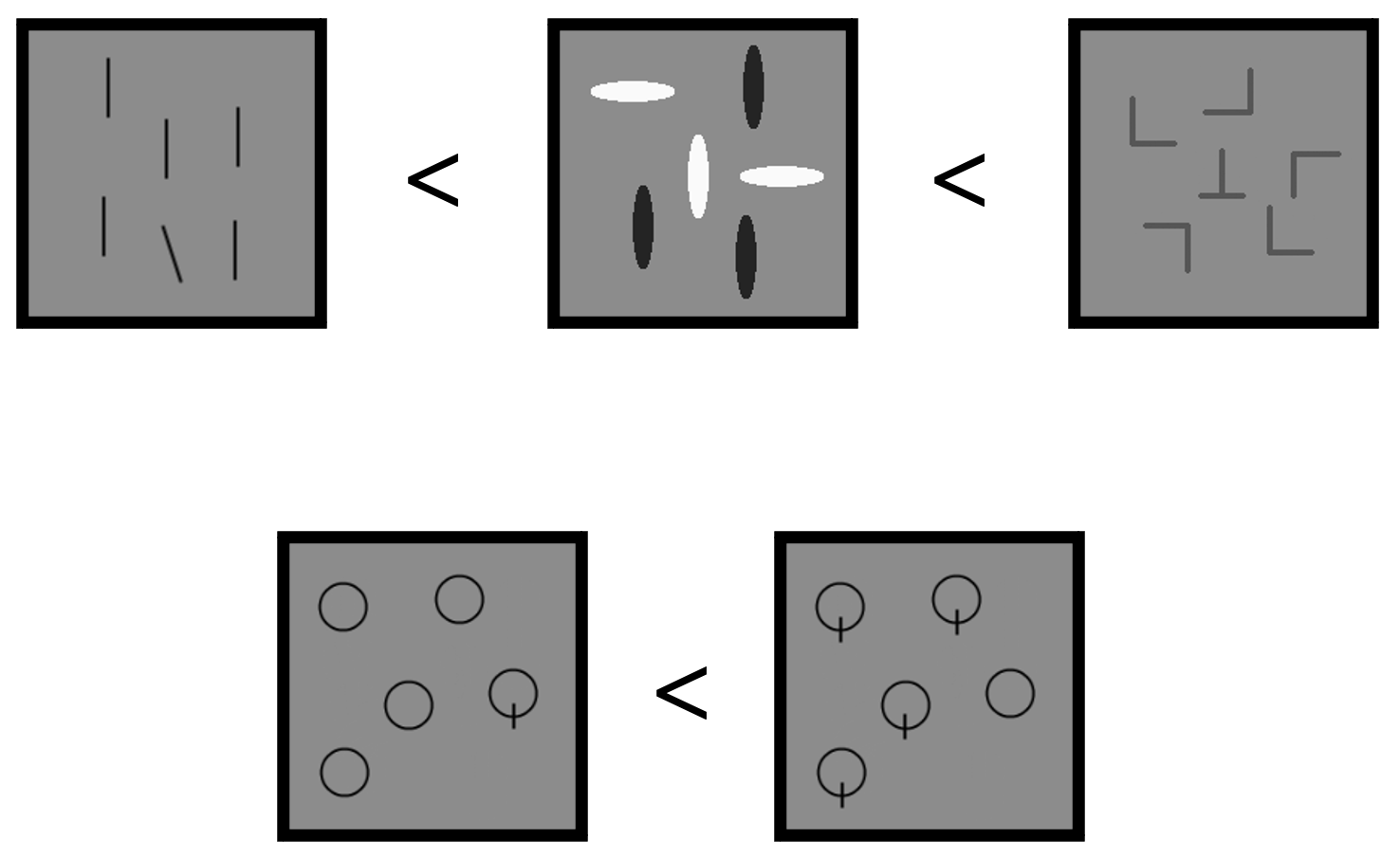}
\caption{Visual search has many peculiarities. Even if one holds constant the discriminability between the target and other search items (the \textit{distractors}), one finds big differences in performance, with search for a unique feature (left) usually far faster than search for a conjunction of features (middle, white AND vertical), which is in turn faster than search for a configuration (right, the two bars configured to form a T). Furthermore asymmetries abound, in which, for instance, it is easier to find a Q among Os than an O among Qs. Such asymmetries are puzzling, given that a Q should be just as discriminable from an O as an O is from a Q.}
\label{fig:search_types}
\end{figure}

\section{Background}
\label{sec:background}

Understanding how human peripheral vision works is important for computer graphics, which often needs to avoid unneeded computation wherever possible; for high-resolution displays it can be very expensive to render in full detail over the entire screen. However, it is not the case that vision in the periphery is simply lower-resolution; under-sampling alone does not account for peripheral loss of acuity.  Over the last few decades vision scientists have explored a number of odd perceptual phenomena in the periphery.  We review some of that work here, focusing on a simple encoding model that explains a surprising amount of the observed behavior.  We demonstrate significant computational improvements to (Section~\ref{sec:model}) and evaluation of (Section~\ref{sec:experiment}) this model.

\subsection{The peculiarity of peripheral vision}

\begin{figure}
\includegraphics[width=0.99\columnwidth]{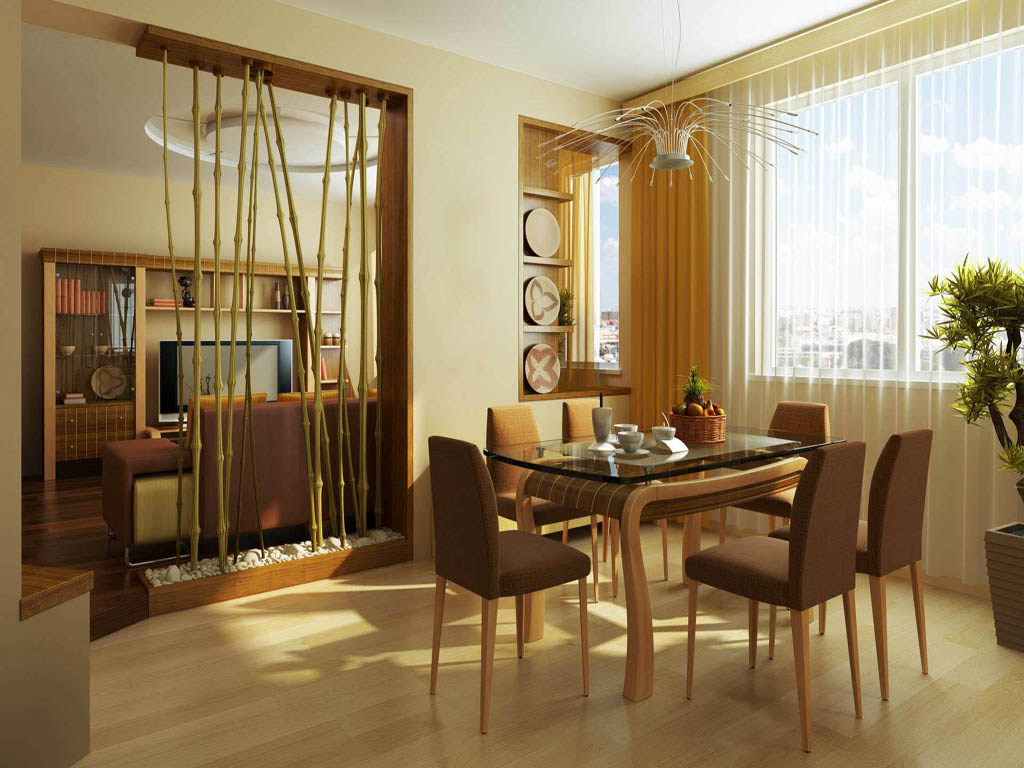}
\includegraphics[width=0.99\columnwidth]{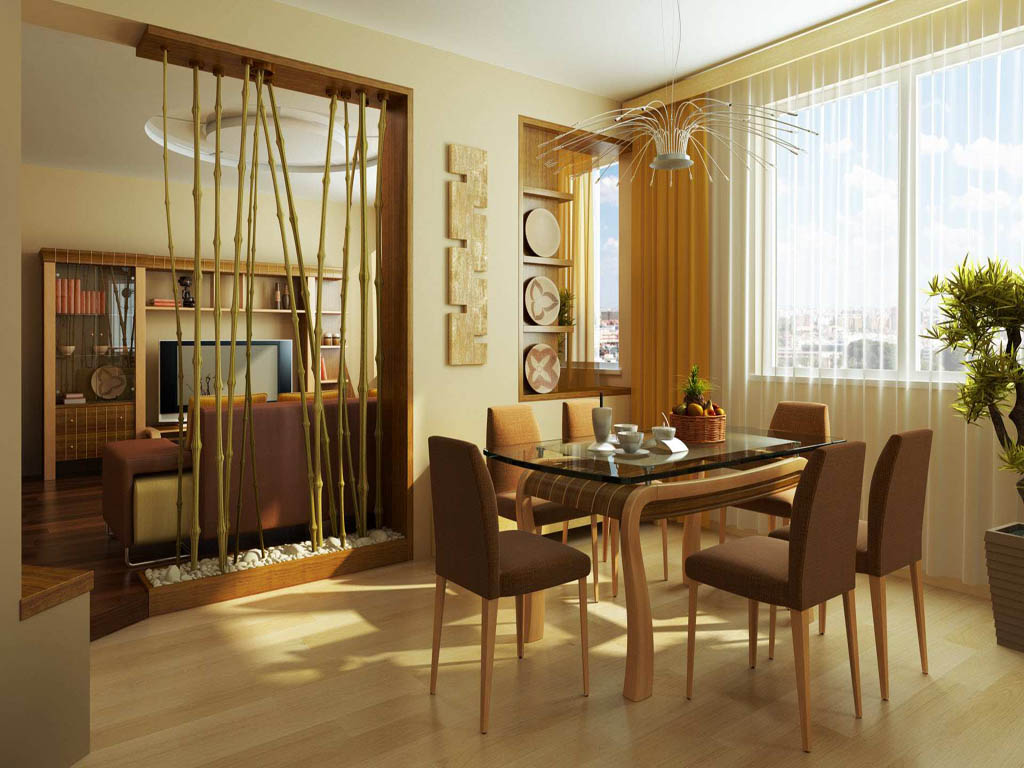}
\caption{Example of change blindness; it is difficult for most viewers to ``spot'' the difference between the two images.  One of the 128 examples available from Sareen et al.~\shortcite{sareen2016cb}}
\label{fig:change_blindness}
\end{figure}

Human vision is not the same everywhere, i.e. across the visual field. We see evidence for this in many visual phenomena. For example, visual search is often difficult~\cite{treisman1980feature}: it can be hard to find your keys, even when they are right in front of you. It is easy to tell your keys from your cell phone; if vision were the same everywhere, search should be easy. That search is difficult implies that vision is not the same across the visual field. The visual search literature is in fact full of such phenomena (Figure~\ref{fig:search_types}): even when it is easy to tell the target from other items in the display, some searches are much harder than others, and there are numerous search asymmetries. A similar conclusion comes from examining human scene perception. A very short glance -- in some cases for as little as 30 ms~\cite{potter1976short}-- suffices to get the gist of a scene. However, if we experimentally probe the details of that percept, we often find that they are murky. One way this can be demonstrated is using a phenomenon known as change blindness, in which observers are asked to find the difference between two similar images, as in a child's puzzle (Figure~\ref{fig:change_blindness}).  Viewers also have trouble with two images presented in succession, and this is essentially the same phenomenon as the side-by-side case~\cite{scott2000comparison}.  The fact that people are often ``blind'' to the differences between the two images~\cite{simons1997change} again suggests that vision is not the same everywhere; if it were, it should be easy to find the differences, as they are easy to see once found. 

%\PS{Maybe reference that two images presented in succession (the typical change blindness demo these days) is the same phenomenon as difficulty spotting the difference in two side-by-side images~\cite{scott2000comparison}}

Both of these phenomena, as well as others, suggest that human visual processing has an information bottleneck. Because of capacity limits, one strategy that the visual system uses is to finely encode information at the fovea, where the observer is \textit{fixating} or pointing their eyes, and more coarsely encode information elsewhere. This strategy degrades information where you are not looking, and at any given moment this means degraded information almost everywhere. The central rod-free fovea subtends only about 1.7 degrees of visual angle, about the size of one's thumbnail at arm's length. Conservatively, 99\% of the visual field lies outside that central foveal region and is subject to a more compressive encoding.

\begin{figure}
\includegraphics[width=0.49\columnwidth]{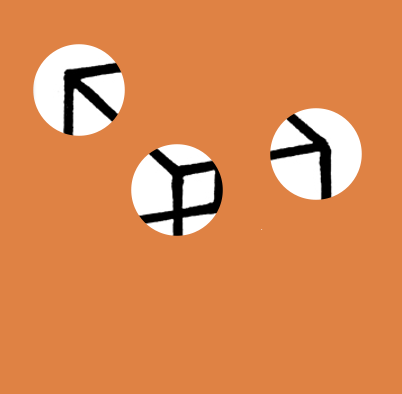}
\includegraphics[width=0.49\columnwidth]{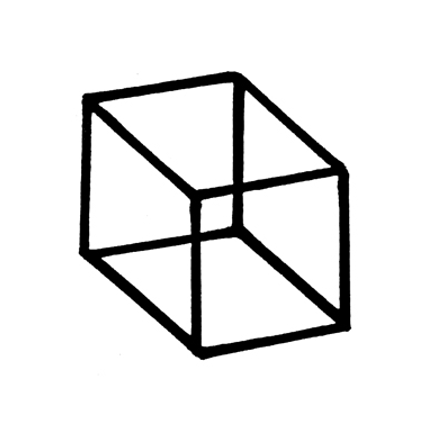}
\caption{If the viewer takes three ``foveal samples'' at ``important'' parts of a wireframe cube it is still hard to gain that the \textit{gist} of the image is a wireframe cube.  This implies that there is important information extracted where the viewer is not ``looking''.}
\label{fig:necker}
\end{figure}

The nature of this encoding, and the information that survives the bottleneck, has significant implications for what tasks a human observer can and cannot perform quickly and easily. First, the information available in the periphery helps the visual system decide where to look next to gather more information. Second, many tasks can be done very quickly and in a single fixation, which suggests that these at-a-glance tasks must rely on information coarsely encoded in peripheral vision. Finally, even though in theory one could make a series of fixations and integrate the information from multiple fixations, in practice humans are not very good at this piecing together of information~\cite{Hochberg1968mindseye}. Researchers have demonstrated this by allowing observers to view an image only through a narrow window~\cite{Hochberg1968mindseye,young2013eye} (Figure~\ref{fig:necker}). Perception under such conditions is not nearly as rich as what one can quickly and easily get from a brief glance at the entire image. Understanding what information is available where one is not looking is therefore critical to understanding vision.
%\PS{Ruth is there a reference for ``researchers have demonstrated'' above?} \RR{Both the Hochberg reference does that, and also the Hulleman reference I've suggested in the past... Young and Hulleman 2013? Hulleman and Olivers? I forget.}

\begin{figure}
\includegraphics[width=0.99\columnwidth]{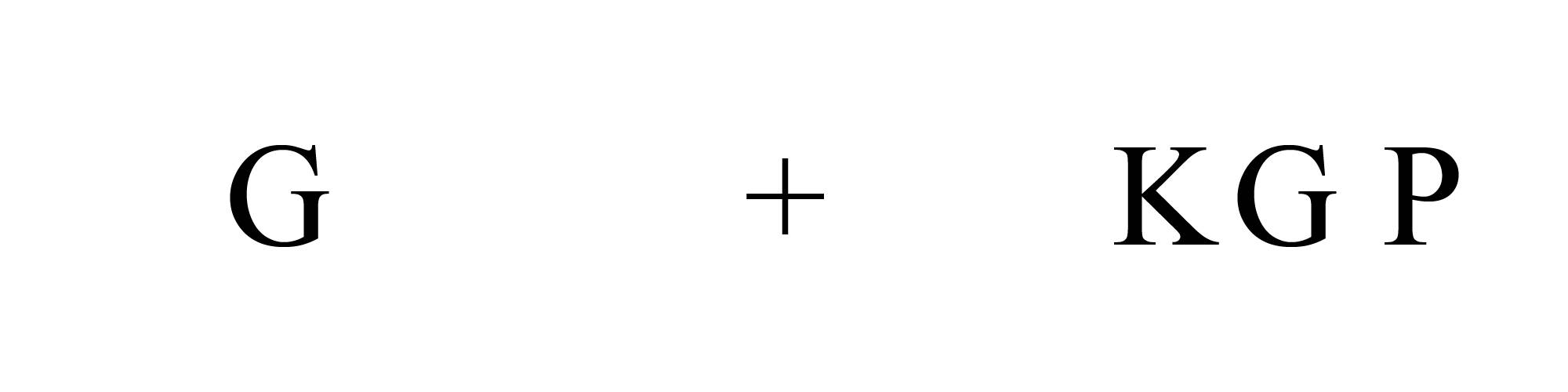}
\includegraphics[width=0.99\columnwidth]{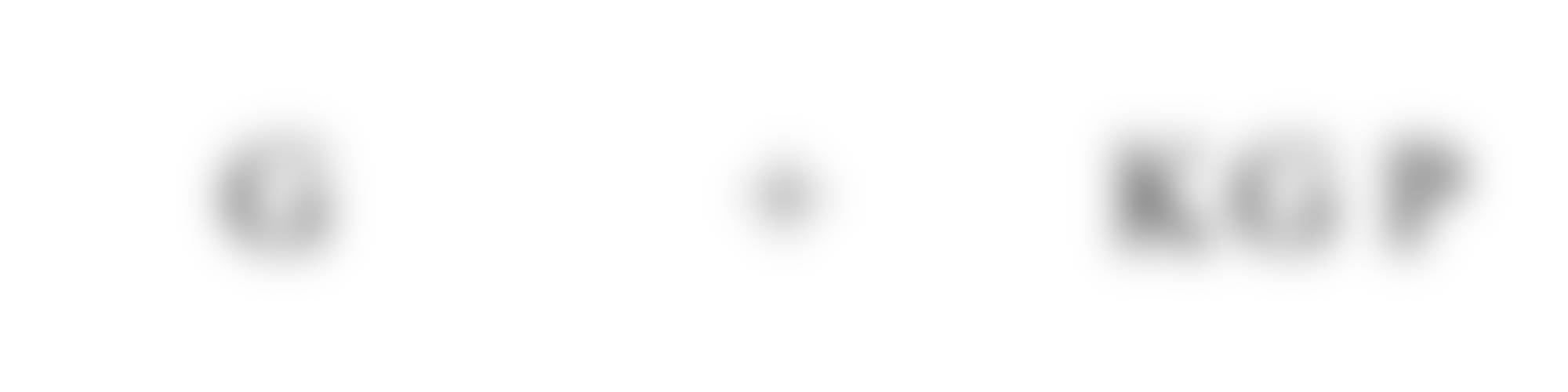}
\includegraphics[width=0.99\columnwidth]{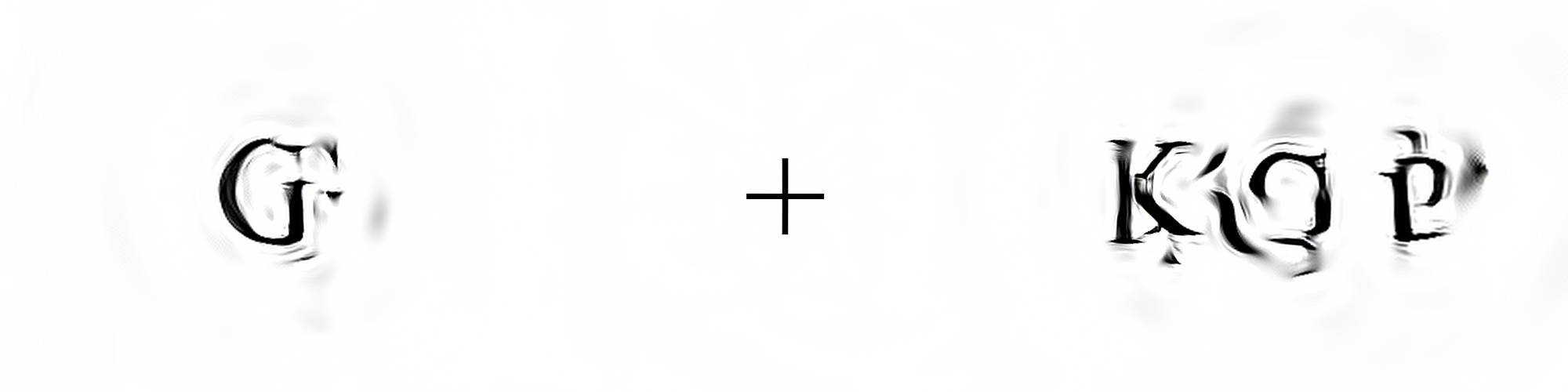}
\caption{(top) Demonstration of visual crowding. Fixating the cross, it is easy to identify the the isolated letter on the left, but hard when that letter is flanked by other nearby letters (adjust your view distance until this is true). (middle)  We can demonstrate that crowding does not merely reflect a lack of peripheral acuity by blurring the image. If the resolution were low enough to interfere with reading the crowded `G' on the right, it would also interfere with reading the isolated `G' on the left. (bottom) Applying our metamer model to the image shows the perceptual asymmetry between crowded and uncrowded letters in the periphery.}
\label{fig:crowding_example}
\end{figure}

Furthermore, vision science research suggests that the compressive encoding used in peripheral vision is not simply a loss of resolution, but is rather quirky and interesting. One often hears that peripheral vision has low acuity, or resolution, relative to the fovea, as peripheral vision more poorly resolves high spatial frequencies. However, it turns out this loss of acuity is actually relatively mild~\cite{rosenholtz2016capabilities}. Furthermore, peripheral vision's loss of acuity cannot explain most change blindness, nor most search difficulty, for which the stimuli have typically been designed to avoid acuity limits. Rather, the big effect in peripheral vision concerns its degradation in the face of clutter, known as crowding. Crowding has often been demonstrated using arrays of letters like that shown in Figure~\ref{fig:crowding_example}. Fixating the central ``+'' at a normal reading distance, one likely has no difficulty reading the isolated ``G'' on the left. Add additional clutter, the flankers ``K'' and ``P'', and it becomes much harder to read the crowded ``G'' on the right. Move those flankers farther from the target letter, and at some ``critical spacing'' the task becomes easy again. We know that crowding is not just due to the loss of peripheral acuity~\cite{levi2009crowding}.  One can easily demonstrate this: if a subject fixates the crowded ``G'' on the right, and we progressively blur the image (Figure~\ref{fig:crowding_example}, middle) until it becomes difficult to read, at that level of blur it is also hard to read the uncrowded ``G'' on the left. Furthermore, examining the blurred image it is obvious that even at this excessively low resolution the flankers do not interfere with perception of the central target. Rather, there must be some other kind of loss of information in peripheral vision. Furthermore, this loss does not apply merely to letter arrays, but rather to perception of any visual stimuli~\cite{rosenholtz2016capabilities}. 

The phenomenology of crowding has given us some hints as to the underlying mechanisms. Under conditions of crowding, firm localization of detail becomes extremely difficult~\cite{korte1923gestaltauffassung,levi1986sampling,bennett1991effects,rentschler1985loss} . Lettvin~\shortcite{lettvin1976seeing} remarked that ``It is not as if these things go out of focus -- but rather it's as if somehow they lose the quality of `form'{}''.  A peripherally viewed word ``only seems to have a `statistical' existence... [preserving] every property save that of the spatial order that would confer shape.'' Peripheral vision tolerates considerable image variation without giving us much sense that something is wrong, despite such variation appearing blatantly obvious when viewed foveally~\cite{freeman2011metamers,koenderink2012space}.
%(NOTE skipping caveats from Wichmann et al, because we're not talking about metamerism, here. Probably should review this in brief later, though.)

\subsection{The summary statistics encoding hypothesis}

These phenomena--the jumbled percept, loss of location information, and the seemingly statistical nature of the perceived stimulus--have pointed a number of researchers toward a particular explanation. Crowding has been attributed to ``excessive or faulty feature integration'', ``compulsory averaging'', or ``forced texture processing''~\cite{balas2009summary,lettvin1976seeing,levi2009crowding,parkes2001compulsory,pelli2008uncrowded}, resulting from an encoding scheme that \textbf{pools summary statistics over local regions}~\cite{balas2009summary,parkes2001compulsory,pelli2004crowding}.

%\RR{I will probably tend to repeat myself in these figure captions vs. the text. Feel free to edit.}
One can make a number of arguments for why the visual system might implement such an encoding. Averaging in the sense of collecting a rich set of \textit{summary statistics} preserves a great deal of useful information, at the cost of discarding precise phase and location of details~\cite{balas2009summary}. Take, for instance, the 700-1000 summary statistics used by Portilla \& Simoncelli's~\shortcite{portilla2000parametric} state-of-the-art model of texture appearance, which includes such summary statistics as the distribution of luminance and color, as well as pairwise cross-correlations applied to the output of oriented filters at multiple scales, similar to early visual processing areas. These latter statistics to some degree encode, for instance, the presence of extended contours, periodic structures, the presence of corners or other junctions, and the sharpness of edges. Portilla \& Simoncelli~\shortcite{portilla2000parametric} used their texture analysis/synthesis procedure to generate new samples of ``texture'' that have approximately the same summary statistics as the original, demonstrating that these summary statistics do a good job of capturing the appearance of textures. This type of statistical textural representation provides an efficient encoding, of use for getting around an information bottleneck in vision~\cite{balas2009summary}.

Furthermore, considerable work in vision science has suggested that many visual tasks are inherently statistical tasks. Texture segmentation and discrimination rely on summary statistics~\cite{rosenholtz2015oxford}. Deviation from local summary statistics makes an unusual item \textit{pop out}, seeming to draw the observer's attention~\cite{rosenholtz1999simple,oliva2003top,gao2007bottom}. Deciding whether a material is shiny might have to do with sub-band skew~\cite{motoyoshi2007image}. Real-world scenes contain many textured regions --- trees, sky, building facades, stone walkways --- and representing those textures well may facilitate scene perception~\cite{oliva2001modeling}. A statistical encoding in peripheral vision, then, may support many real-world tasks.

\begin{figure*}[hbt!]
\includegraphics[width=\textwidth]{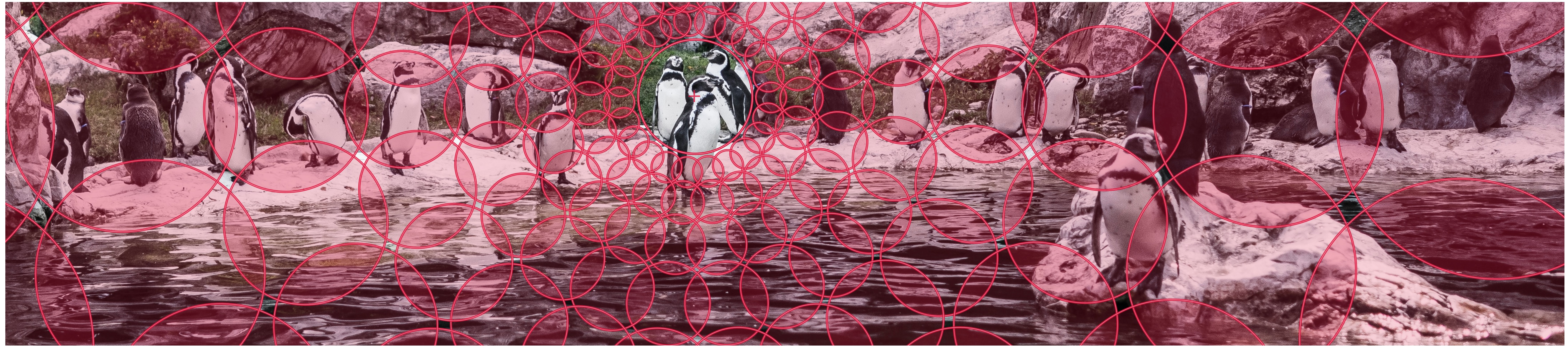}
\caption{Visualization of a partial lattice of foveated pooling regions overlaid on an image, with a viewing position centered on the middle three penguins.  As one moves from the center of fixation into peripheral vision, pooling regions become progressively larger, integrating information over larger regions, such that at the edges of the image multiple objects fit within an entire pooling region. For visualization clarity, the pooling regions shown are sparser than those used to create our metamers, and the central 5 degree foveal region has been shown without pooling.}
\label{fig:pooling_regions}
\end{figure*}

\begin{figure*}
\includegraphics[width=\textwidth]{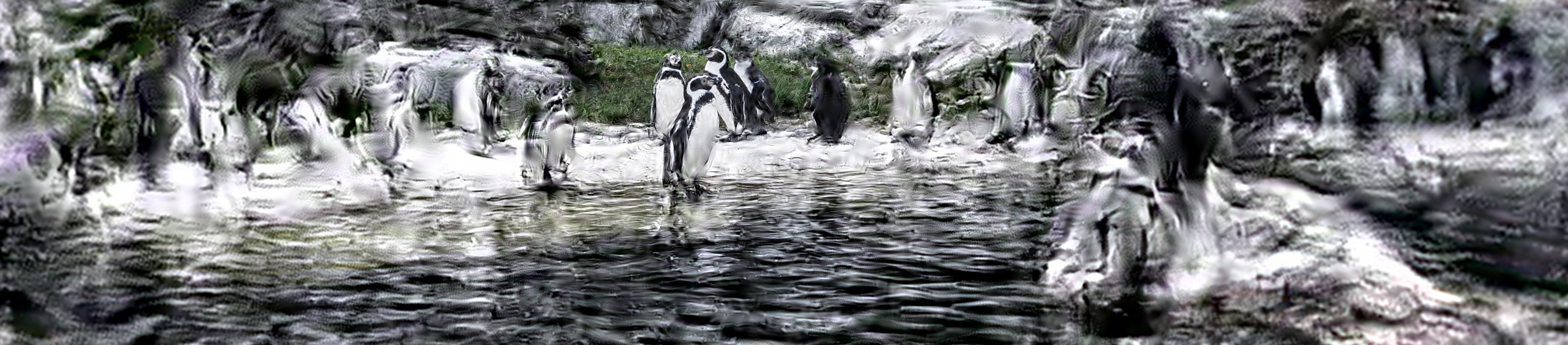}
\caption{A metamer of the same image used in Figure \ref{fig:pooling_regions} generated with our peripheral encoding model using the same fixation point. When centering gaze at this fixation, the summary statistics under an ideal model should be the same as in the original within each pooling region, and the distortions at the image edges would be undetectable in peripheral view. This is despite the extremely distorted appearance of the image away from the modeled fixation, when viewing those regions at the center of gaze.}
\label{fig:penguin_met}
\end{figure*}

%\RB{Need stitching text between these two paragraphs -- end this section by reiterating why all these perceptual limitations are important for graphics}
% Pete will try to do this with just the first sentence
A peripheral encoding scheme in terms of pooling of summary statistics over sizeable regions is good news from the point of view of applications like image or video compression, and rendering in computer graphics. Because peripheral vision loses substantial information with this scheme, it will be less aware of and less able to discriminate compression or rendering artifacts and some kinds of distortions and details. However, we do not completely ignore our periphery. We use peripheral vision for a number of tasks, including detection of interesting or salient targets; observers will notice losses due to rendering shortcuts, video compression, and so on, if we do not adequately account for the strengths and limitations of peripheral vision.

%Limitations of peripheral vision are good news from the point of view of applications like image or video compression, and rendering in computer graphics. Because peripheral vision loses information, it will be less aware of and less able to discriminate compression or rendering artifacts and some kinds of distortions and details. However, we do not completely ignore our periphery. We use peripheral vision for a number of tasks, including detection of interesting or salient targets; observers will notice losses due to rendering shortcuts, video compression, and so on, if we do not adequately account for the strengths and limitations of peripheral vision.

\subsection{Image Synthesis From Summary Statistics}

To model peripheral vision, one needs first to capture the loss of information in peripheral vision, followed by an eccentricity-dependent version of pooling (Figure~\ref{fig:pooling_regions})~\cite{freeman2011metamers,rosenholtz2011your}. This involves computing summary statistics within each of multiple pooling regions that grow linearly with \textit{eccentricity} (i.e.\ distance to the point of fixation), overlap, and densely tile the visual field. Such an encoding, based on a rich set of summary statistics computed over local, eccentricity-dependent pooling regions, is the state-of-the-art model of peripheral vision. The loss of information in the hypothesized peripheral encoding can predict difficulty recognizing peripheral objects in cluttered displays or scenes~\cite{balas2009summary,chang2016search,freeman2011metamers,keshvari2016pooling,rosenholtz2012summary,zhang2015cube}. The loss of information also predicts difficult search conditions, while preserving the information necessary to predict easy search~\cite{rosenholtz2012summary,chang2016search,zhang2015cube}. In spite of this loss of information, the encoding preserves sufficient information to predict the ease with which observers get the gist of a scene at a glance, including identifying the scene category, upcoming turns when driving, and the presence of target objects like an animal or a stop sign~\cite{rosenholtz2012rethinking,ehinger2016general}.  

As in the demonstration in Figure~\ref{fig:crowding_example}, one can gain intuitions about what information is preserved and lost in peripheral vision by synthesizing images with approximately the same summary statistics. Of relevance to the present paper, synthesizing these images (such as the one shown in Figure~\ref{fig:penguin_met}) based on the constraints from multiple pooling regions can pose a significant computational challenge.  Portilla \& Simoncelli~\shortcite{portilla2000parametric} pool these summary statistics over a single pooling region covering the entire input image, and synthesizes a new texture by starting with a random image, and iteratively applying the summary statistics as constraints, adjusting the image pixels until the synthesized image converges to have approximately the same statistics as those measured in the original. CNNs have also been used to synthesize new textures with approximately the same summary statistics, for different sets of statistics. For example, \cite{gatys2015texture,wallis2017parametric} used the intermediate activations of a pre-trained CNN (VGG-19), pooled over the entire image, as their statistics.  Compared to Portilla \& Simoncelli~\shortcite{portilla2000parametric}, they achieve much better quality when viewed foveally but roughly equal quality when viewed peripherally and use roughly ten to hundred times more statistics so are more computationally intensive. However, the use of these summary statistics derived from a trained CNN is far less well-tested as a model of peripheral vision, per se.

Directly extending Portilla \& Simoncelli~\shortcite{portilla2000parametric} to multiple regions, using a related Fourier approach, as done by Freeman \& Simoncelli~\shortcite{freeman2011metamers},
takes on the order of 6 hours to synthesize one image from a 512x512 original.  We use a different approach, aimed at efficiency and flexibility, which directly runs a data flow network with the desired statistics wired into it (rather than being learned or computed in a Fourier space) as detailed in the next section.

\section{Perceptual Encoding Model}
\label{sec:model}

In the peripheral encoding model an image is represented by a set of image statistics that have been averaged over local pooling regions.  If two images have matching pooled statistics, then the model predicts they will be visual metamers. Loosely speaking, the statistics encode the presence of local ``features'', especially edge-like features, while the pooling removes information about the precise location of these features. Thus two metamers will contain similar features but potentially rearranged or jumbled within the pooling regions. This approach to peripheral perception was pioneered in the work of \cite{balas2009summary,freeman2011metamers,rosenholtz2012rethinking}.  

In our implementation, the pooled statistics are computed using three stages.  
First the input is converted into a set of component images split by color channel, spatial scale, and orientation. The component images are based on an augmented form of Steerable Pyramids \cite{freeman1991steerable}.
Next we compute statistic images which correspond to various moments and correlations of the component images.  As with prior methods, our statistics are based on the texture synthesis work of Portilla \& Simoncelli~\shortcite{portilla2000parametric}, though we use a modified the set of statistics, as we will discuss later.  
Then each statistic is locally averaged to compute a single value for each pooling region, effectively blurring and downsampling the statistic images.  The result is stored as a stack of low-resolution pooled images, one for each statistic.  Typically we use approximately 300 statistics for grayscale images and roughly one thousand for color images. For comparison, Freeman \& Simoncelli~\shortcite{freeman2011metamers} and Rosenholtz et al.~\shortcite{rosenholtz2012rethinking} use between 700 and 1000 statistics for grayscale images.

To create metamer images, we first compute the pooled statistics for a target image.  Then we iteratively adjust a seed image using a gradient-based optimization process until its pooled statistics closely match those of the target image. Our system is implemented using the PyTorch library \cite{PyTorch2019} which provides automatic gradient computation, highly optimized image operations, and easy ability to utilize GPUs. This has allowed us to work with higher resolution images and greater freedom to experiment with different statistics as compared to prior work.  
Using automatic differentiation allows us to easily modify our statistics without having to invent new optimization routines.  Our system's structure resembles a convolutional neural network, though in our case, the network is fixed and we learn an input to match a desired output.
We plan to release our system as open source.

\subsection{Component Images}

For color images we first split the image into color channels. We use a perceptually-based three-channel opponent color space, which consists of achromatic, red-green, and blue-yellow channels~\cite{CIELAB}. For grayscale images only the achromatic channel is used.  Each channel is then handled mostly independently and symmetrically in our system though we do include a few cross-channel color correlation statistics.  There is relatively little discussion of color in the prior work.  Freeman \& Simoncelli's \shortcite{freeman2011metamers} sample code uses a custom per-image color space based on a principal components analysis (PCA) while Rosenholtz~\shortcite{rosenholtz2011your} used independent components analysis (ICA).  However these per-image color spaces are content-dependent and harder to extend beyond still images. We found that using opponent color channels is simpler and works well across a variety of images.

\begin{figure}
\includegraphics[width=0.69\columnwidth]{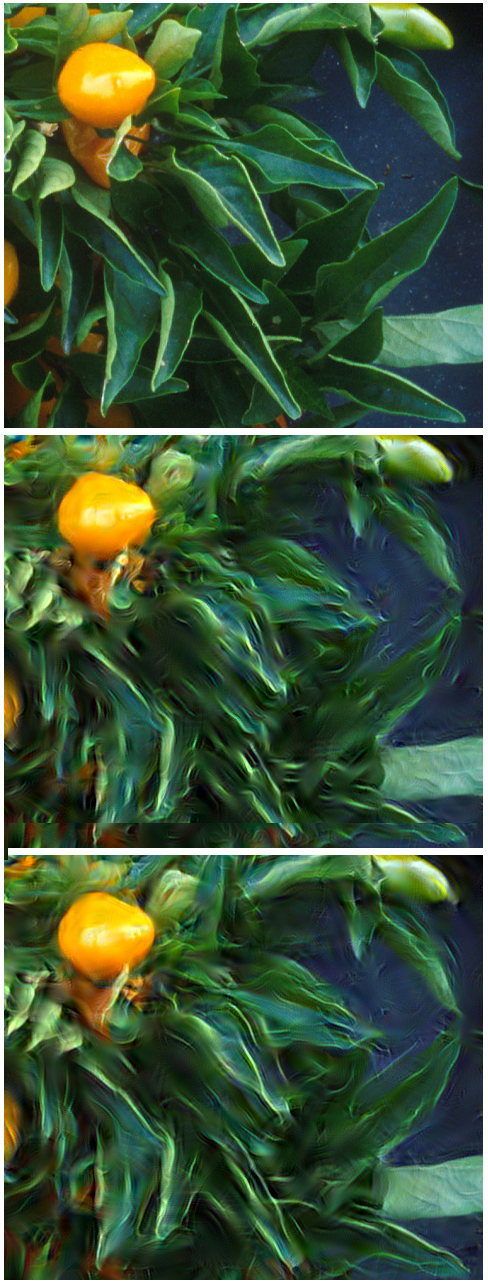}
\caption{Including two additional filter orientations and end-stopping statistics noticeably improves the quality of generated metamers. Shown are example insets from the original (top), 4-orientation/non-end-stopped metamer (middle), and metamer generated with 6-orientations and end-stopped statistics (bottom). These two additions greatly improve the ability of the model to reproduce high-contrast, continuous, curved edges. Note that while these statistics improve the leafy section of the image, they have little effect on the yellow pepper in the top left. Image from \cite{usdapeppers}.}
\label{fig:peppers}
\end{figure}

Next, for each channel we construct an augmented Steerable Pyramid in the same style as in Portilla \& Simoncelli ~\shortcite{portilla2000parametric}. A steerable pyramid uses a set of convolutional filters tuned to different spatial scales and orientations to split the image into corresponding components. Most of these filters resemble Gabor-filters and are very good at detecting edge-like features in the images.  Steerable pyramids can be viewed as a soft partitioning of the image in Fourier space and can be computed using Fourier transforms for efficiency.  The augmented steerable pyramids contain a few extra components.  Each edge filter is replaced with a quadrature pair (filters offset by a 90 degree phase shift) \cite{freeman1991steerable} to better track edge polarity and correlations.  The quadrature pairs are also combined to create edge magnitude images.  In addition Portilla \& Simoncelli~\shortcite{portilla2000parametric} added phase-double images for tracking cross-scale correlations.  We use the same components as Portilla \& Simoncelli~\shortcite{portilla2000parametric}, except that we have extended the pyramids to use 6 spatial levels instead of 4 and use 6 orientations instead of 4 (Figure~\ref{fig:peppers}).  These changes improved our ability to create metamers for images with features at a wider range of scales and for images with strongly directional patterns.

\subsection{Image Statistics}

The image statistics correspond to various moments and correlations of the component images.  Our statistics are based on those of Portilla \& Simoncelli~\shortcite{portilla2000parametric}, though with some significant changes.  For computational simplicity we use raw rather than central statistics (i.e.\ we omit the mean subtraction), as the mean values are also included among the statistics to be matched.
We also removed some statistic types that did not seem to have much effect on metamer quality in our preliminary tests.  In particular we removed the low-pass component statistics and all of the autocorrelation statistics.  This is a considerable savings as numerically these actually constituted a majority of the original statistics.  The remaining statistics for each color channel image include: pixel mean value, edge magnitude mean and variance at each scale and orientation, correlations of edge magnitude across orientations at the same scale, and correlations of edge magnitude and phase between neighboring scales at the same orientation.

For color images, we add a few cross-color-channel correlations, namely, between edge magnitudes and between phase components at the same scale and orientation.  These are important for reproducing multi-channel hues such as orange or purple.
The use of opponent color channels along with these few extra statistics has worked quite well in our tests across a wide range of input color images.

Nearly all our statistics are edge-based and respond most strongly to linear edge features. This corresponds roughly to how simple and complex neurons work in human visual cortex \cite{tn1959receptive}.  The early visual system has other neurons though, such as
hypercomplex \cite{hubel1965receptive} or end-stopped neurons, that respond most strongly only in locations where edges end, curve, or change direction.  End-stopping has been shown to be important in image perception \cite{heitger1992simulation,zetzsche1993importance,julesz1981textons} but is not well represented in the prior statistics.  Thus we have added new statistics that better capture end-stopping by measuring the change in strength along an edge.  It is computed by subtracting each edge magnitude component image from a copy of itself translated a short distance along the expected edge direction (i.e. the filter orientation) and then squaring the result.  In our tests, these new end-stop statistics improved metamer quality in regions with lines that curve or end (Figure~\ref{fig:peppers}).

\subsection{Pooling Regions}

Each image is conceptually subdivided into set of overlapping pooling regions.  Within each pooling region only the average value of each statistic is kept.  
This is intended to mimic the way visual inputs are processed and pooled within our visual cortex.  The size of such pooling regions varies with angular distance from the gaze point in an approximately linear relationship sometimes called Bouma's Law.  Thus the maximum acceptable metamer pooling region size depends on the eccentricity, or visual angle from the point of gaze.  

Our system can generate two different patterns for the pooling regions: uniform and gaze-centric.  Uniform metamers use the same size pooling region everywhere and are thus only suitable for peripheral viewing anywhere beyond a fixed target eccentricity.  Whereas in gaze-centric metamers, the pooling size varies linearly with the eccentricity and is thus relative to a specific gaze point.  For efficiency we compute gaze-centric metamers by first warping the image into a log-polar space to equalize the pooling region sizes and shapes, computing a uniform metamer in this space, and then transforming the result back to normal image space.  The axes of the log-polar space are the logarithm of the radius from the gaze point and the azimuthal angle around the gaze point with scaling factors chosen such that all the transformed pooling regions are equal-sized circular regions of a chosen size.

In our system the pooling is computed by convolving the statistic images with a pooling kernel, effectively blurring them, and then downsampling the result to reflect the limited number of pooling regions.  The results are low-resolution pooled statistic images where each pixel represents one pooled statistic in one pooling region.  Our pooling kernels are based on those used in in Freeman \& Simoncelli~\shortcite{freeman2011metamers} with a few significant changes.
We use circular pooling regions rather than elliptical ones, though sized to have approximately the same area as in their V2 model.  We also use a higher density of pooling regions with much more overlap between them.  The spacing between neighboring pooling regions in their system was 3/4 of its diameter but only 1/4 of the diameter in ours.  Since we do not know where a viewer's actual pooling region boundaries will be, we used this high density to ensure there will be a closely-aligned computation pooling region for any potential viewer's actual pooling regions.  However our density is likely higher than actually needed and we plan to try reducing it in future work. 

The supplemental material includes a video showing applications of our model to dynamic viewing conditions, including a static image with changing gaze location, an example of the convergence of our iterative solver, and a short dynamic movie with static gaze.

\begin{figure*}[hbt!]
\centering
\includegraphics[width=1.0\textwidth]{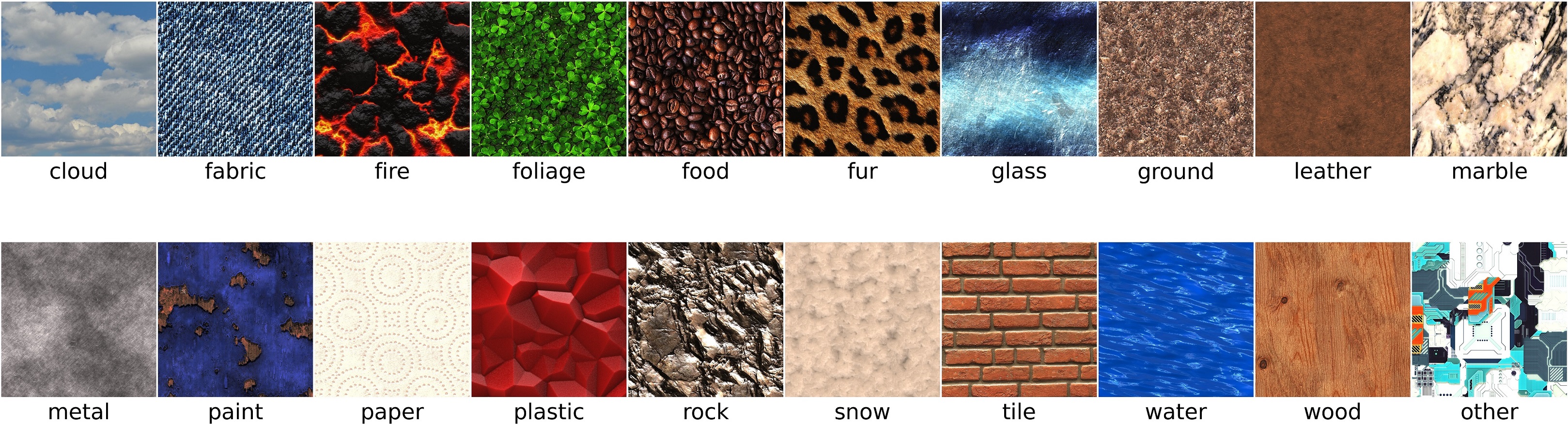}
\caption{Our set of 400 textures spans a range of texture material quality descriptor categories. A single exemplar texture for each of the 20 texture material categories is shown. Contrast enhanced for better viewing at a small scale.}
\label{fig:material_categories}
\end{figure*}

\section{Experimental Evaluation}
\label{sec:experiment}

Our new dataflow model enables experiments we could not easily run before. Many experiments require generating a lot of metamers. For example, if one wanted to test whether a peripheral encoding model could predict scene perception, or if one wanted to test how much these syntheses produced scene metamers indistinguishable from the originals, one needs to test a lot of scenes. As a result, one might need to synthesize hundreds of images. Synthesizing 400 640x480 grayscale images would take as much as 6 hrs x 400 = 100 days. This is a big commitment to run a simple experiment! As a result, a fair amount of previous work has used a very small number of images. For example, while Ehinger \& Rosenholtz~\shortcite{ehinger2016general} used a cluster of computers to synthesize 400 grayscale images in a ``mere'' week, to test whether the information in peripheral vision could explain difficulty performing scene perception tasks, Freeman \& Simoncelli~\shortcite{freeman2011metamers} tested scene metamerism (whether you could tell one metamer from another while fixating) with a mere 4 images, and Wallis et al.~\shortcite{wallis2019image} with a mere 20 images. Alternatively, some researchers have used a CNN-based model of similar architecture to the summary statistic model to synthesize large numbers of model-metamers, e.g.~\cite{wallis2019image}. This has the advantage of computational efficiency, but the disadvantage that, due to dramatic differences in the early stages of the models, we do not know whether such CNN-based models actually model peripheral vision.

With our new dataflow model, we can synthesize 400 full color 1920x1080 images on a single GPU in a more manageable 23 hours, facilitating many experiments. Here we use this capability to study metamerism in texture perception, which is relevant in many graphics applications. Video games often use variable level of detail for both scene geometry and object textures (i.e. mip-mapping) in distant regions away from likely fixation, and some foveated rendering techniques such as fCPS~\cite{patney2016towards} explicitly use texture downsampling in the periphery. Even within a complex scene, perception of certain objects (i.e. trees) can be thought of as texture perception in terms of the image that is projected onto the eye \cite{oliva2001modeling} indicating that our findings should have some application for metamerized scenes as well. 

Previous work has examined whether or under what conditions participants can distinguish between original and synthesized textures shown in the periphery. Much of this work was restricted to texture perception in the near-periphery, or collected statistics in a non-foveated way, over the entire texture patch, using the Portilla \& Simoncelli~\shortcite{portilla2000parametric} texture synthesis model. The advantage of not using a foveated model is speed (about 2 minutes to synthesize a 256x256 patch), but makes no distinction between foveal and peripheral perception, discarding the same information regardless of eccentricity of the patch. This disadvantages non-homogeneous textures and those with larger scale structures. 

Balas~\shortcite{balas2006texture} examined 15 textures from the Brodatz database \cite{brodatz1966textures}, and examined which statistics measured in Portilla \& Simoncelli~\shortcite{portilla2000parametric} were necessary and sufficient for capturing appearance for those textures. Textures as displayed to the participants extended from near the fovea out into the periphery; due to the statistics being pooled over the entire image, artifacts in or near the fovea may have allowed participants to distinguish between original and synthesized textures. Keshvari and Wijntjes~\shortcite{keshvari2016peripheral} studied material identification in (1) a 2 deg diameter patch of original texture at fixation, (2) a 2 deg diameter \textit{metamer} synthesized using Portilla \& Simoncelli statistics measured over the image as a whole, also shown at fixation, or (3) a 2 deg diameter patch of original texture at 10 deg eccentricity. Their test set included 50 images in each of 6 material categories: fabric, foliage, stone, water, and wood. They found that material perception with the synthesized images predicted peripheral but not foveal material perception with the original images. Finally, Wallis et al. \cite{wallis2019image} examined the foveated encoding of Freeman \& Simoncelli~\shortcite{freeman2011metamers} (pooling of statistics over multiple overlapping pooling regions), for 10 ``texture-like'' images, and found that observers had difficulty telling apart original textures from synthetic.  They found people could reliably distinguish some images from their Freeman \& Simoncelli~\shortcite{portilla2000parametric} metamers and concluded that a pooling model is insufficient and content-dependent grouping and segmentation would be needed.  However, it could also be that the Freeman \& Simoncelli~\shortcite{portilla2000parametric} statistics are incomplete.

\subsubsection{Our Approach}
Our experimental design differs significantly from the design used in many previous related works. This choice to break with tradition was motivated by several factors. First, we leverage the speed of our system to synthesize and present high resolution color stimuli with more than twice the field of view of any previous study on metamers (29 deg vs 25 deg) \cite{wallis2019image}. Additionally, we increased the stimulus duration from 300 ms to 2.5 sec and presented stimuli side-by-side rather than sequentially in time. These changes require that the model be correct at a much broader range of eccentricities and for a longer period of sustained peripheral attention, setting a more stringent (and arguably more realistic) criteria for a successful metamer. Simultaneous presentation of both metamer and original ensures that the texture boundary between the two is also unnoticeable, and provides a better setup for performing the same experiment with video.

Additionally, to study participants' ability to distinguish metamer textures from original, we curated our own data set of 400 texture images (described below). All metamer models (including ours) perform better for some images than for others. Examining a wide variety of textures allows us to begin to assess what image characteristics determine whether a metamer succeeds, i.e. is indistinguishable from the original given a specific viewing distance and direction. Differences in metamer quality did not seem to correlate with the residual loss of the optimization procedure, suggesting that at least some statistical properties of the images that participants can perceive in peripheral vision were not captured by the model. Our goal therefore was to test many different textures to determine which textural qualities might not be well-represented by our statistical descriptors. We also analyze our results in terms of material category, to get additional intuitions.

\section{Rendered Texture Data Set}
\label{sec:data}

We curated a data set of 190 physically-based rendering (PBR) textures that cover a range of material categories: fabric, fire, foliage, food, fur, glass-like, ground-like, leather, marbled, metal, painted, paper, plastic, rock, snow- or smoke-like, tiled, water, wood, and an additional `other' category for interesting textures that did not fit into any of the material groups above. The textures were hand curated across several online databases of high quality free-use PBR materials\footnotemark,\footnotetext{Texture material databases: \cite{3DTexturesdotme}\cite{cc0textures}\cite{patternpanda}\cite{publicdomaintextures}\cite{freepbr}\cite{gumroad}\cite{cgbookcase}\cite{sharetextures}\cite{pixelfurnace}\cite{texturecan}\cite{pbrtexture}\cite{sketchuptextureclub}\cite{texturebox}\cite{texturehaven}\cite{cadhatch}} with the goal of covering both the most commonly used material categories and a representative sample of scale and shape within each category. We sought to both span the range of natural textures, and include texture materials and qualities relevant to graphic artists. The material categories chosen (Figure~\ref{fig:material_categories}) represented the union of previous texture material databases \cite{sharan2009material}, and online texture databases which characterize materials by other qualitative features. 

Each texture was rendered using G3D \cite{G3D17} at half native resolution tiled onto a 1920x1080 RGB output image. Textures were rendered from a fronto-parallel angle under two different cube map lighting environments (white room and plain sky) coupled with with two different angles of directional lighting (oblique and direct, respectively). We also included 20 high resolution photographs of clouds (from \cite{cloudphoto} and the authors' personal collections), since clouds are commonly used as background textures in many rendered outdoor scenes. The total size of the data set was therefore 400 images, 20 images per category across 20 categories.

%In addition to representation over the 18 material categories, we also designed our texture data set for representation over the 47 texture descriptors outlined in \cite{cimpoi14describing}. As opposed to describing the material types present in a given texture image, these texture descriptors correspond instead to textural qualities that are somewhat independent of material such as bumpy, dotted, lined, smeared, etc. While a single material category cannot span all 47 texture descriptors, we chose textures such that over the entire set of N texture images, all texture descriptors were represented. (histogram figure of distribution of textures among 47 or reduced set of texture descriptor categories)

\subsection{Categorical Texture Descriptors}
\label{sec:descriptors}

The size and diversity of our texture data set represents a broad swath of realistic textures, but analyzing the effectiveness of our model for each texture individually would be unwieldy and prevent generalization to new textures. We therefore sought to find unifying textural descriptors that would provide graphics users of this model an intuitive understanding of which textures make good metamers and which do not. Unfortunately, there have been many efforts to develop automated texture classification methods \cite{cimpoi14describing} and the academic community lacks consensus on which intuitive categorical texture descriptors are the right ones to use. Based on multidimensional scaling of human judgments, Rao \& Lohse \shortcite{rao1996towards} proposed three dimensions: repetitive vs. non-repetitive; high contrast non-directional vs. low contrast directional; and coarse low complexity vs. non-granular, fine, and high complexity. However, Liu \& Picard \shortcite{liu1996periodicity} suggested a different, simpler set of three texture descriptors: periodicity, directionality, and randomness. Another set of seven descriptors was proposed by Tamura et al.~\shortcite{tamura1978textural} and are still widely used to characterize images and textures \cite{cusano2021t1k+}. Their proposed texture characteristics are coarseness, contrast, directionality, line-likeness, roughness, and regularity.

\begin{figure}
\centering
\includegraphics[width=0.9\columnwidth]{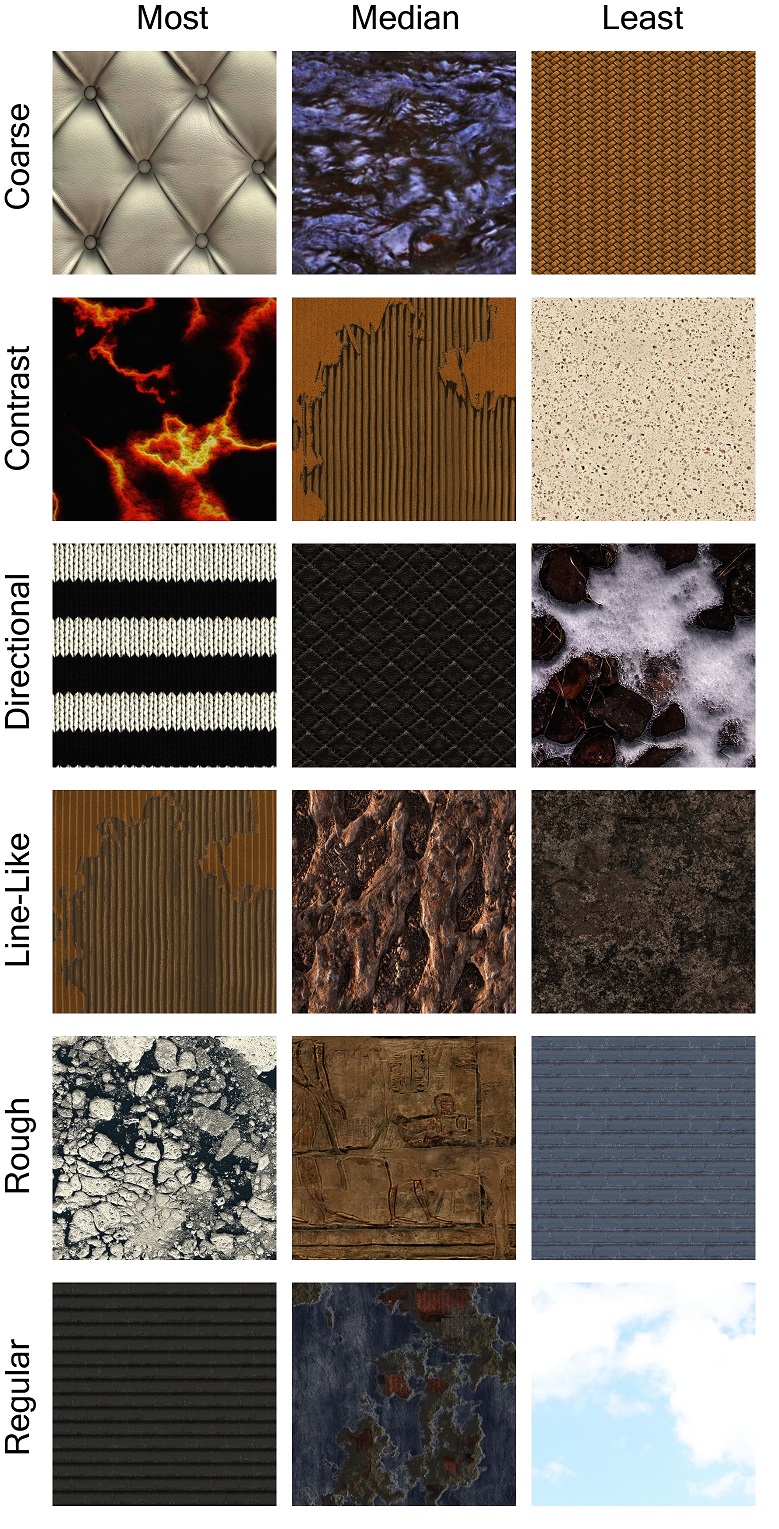}
\caption{The texture features based on \cite{tamura1978textural} match perceptual features in our rendered textures. Shown are the lowest, median, and highest scoring textures for each feature descriptor. Contrast enhanced for better viewing at a small scale.}
\label{fig:texture_features}
\end{figure}

We opted to follow the general direction of Tamura et al.~\shortcite{tamura1978textural}, but updated them to reflect modern practices in both graphics and vision science, and to fix issues they themselves identified with some of their descriptors. In this paper we propose a new set of computational measures that we believe match the intuitive definitions of the original seven descriptors from their 1978 paper. The maximum, median, and minimum rated textures for each descriptor are shown in Figure~\ref{fig:texture_features}. We offer these updated texture descriptors as a contribution to the community and will release our code as open source.

As in Tamura's texture metrics, we sought to represent the textures as close to that seen on the screen by our human subjects and therefore did not normalize their values. Tamura's texture metrics are also defined for grayscale images, whereas our textures were in color; we therefore converted our RGB images to luminance images ($L$) using the equation: $L = 0.2126R+0.7152G+0.0722B$ for linear values of R,G, and B \cite{iec199961966}.

\subsubsection{Coarseness} is a measure of the general scale over which a texture repeats. Tamura et al.~\shortcite{tamura1978textural} measured this by taking the difference in local mean (rectangular filter) both above and below and to the left and right of each pixel at a several scales. Coarseness was then defined as the average (over the whole image) of the maximum response to either of these measures at each location. Taking a more modern approach, we instead filtered using a 2-orientation 6-scale steerable pyramid \cite{simoncelli1995steerable}. This has the advantage that it will find the maximum response at any orientation, not just the horizontal and vertical. The steerable filter images $L$ are defined by:
\begin{align*}
L_{vs} = I \circledast \mathcal{F}_{vs}\\
L_{hs} = I \circledast \mathcal{F}_{hs}
\end{align*}
where filter set $L$ is the convolution with Image $I$ and a filter $\mathcal{F}$, resulting in a filter image for each combination of scale $s$, and orientation (here only vertical $v$ and horizontal $h$). We then followed the same procedure as in Tamura's equation, taking the scale of the maximum for each pixel across all scales, and exponentiating this value - using base 10 instead of base 2 for a better fit to our data set. We applied these exponential/log and range normalizations for many of the statistics, to better allow for combining them together in our estimation of Regularity. Coarseness $Cor$ is then the mean coarseness over the entire image normalized over all textures in the data set such that the values range between [0,1].

\subsubsection{Contrast} measures the distribution of luminance values over an image, typically computing some measure of the variability in luminance relative to the average. Tamura et al.~\shortcite{tamura1978textural} noted that simple contrast definitions do not correlate with psychophysical measurement, and instead used the pixel standard deviation divided by the kurtosis to the $\frac{1}{4}$ power, where the $\frac{1}{4}$ was determined empirically to work best. The pixel standard deviation represents the variability in pixel values, while the kurtosis represents the polarity of the pixels. In our case, we preferred a contrast measure adapted specifically for naturalistic images, and which would take into account local differences in luminance relative to local mean luminance, at different scales.  We used the complex image contrast measure adapted from \cite{peli1990contrast}, based on a Laplacian pyramid decomposition \cite{burt1987laplacian}. The Laplacian pyramid allows one to easily construct a multi-scale representation, $R_{s}$, of the original image, as well as bandpass images $L_{s}$ representing the difference between neighboring scales $(R_{s}, R_{s-1})$. The bandpass image $L_{s}$ effectively gives us the local difference in luminance at a given scale, $s$, whereas $R_{s-1}$ gives us the local mean luminance at scale $s$. Therefore a reasonable measure of local contrast at a given scale $s$ is:

\begin{align*}
Con_{sxy} = \bigg|\frac{L_s}{R_{s-1} + \epsilon} \bigg|
\end{align*}
where the small $\epsilon$ in the denominator avoids division by zero.
Finally, we average contrast over all $xy$ pixels, and over all scales $s$.
\begin{align*}
Con = \frac{1}{sxy} \sum_{sxy} Con_{sxy}\
\end{align*}
We again normalize this mean contrast value by the maximum contrast over all textures, scaling the range from [0,1].

\subsubsection{Directionality} is a metric designed to measure, over the entire image, any biases in overall orientation. For directionality, Tamura et al. first applied a crude 3x3 edge filter in both the horizontal and vertical directions, and measured the local direction, theta, as the arctan of these responses. They formed a histogram of those angles, for all responses for which the $[\Delta X, \Delta Y]$ vector was above some threshold length, and computed the ``peakedness'' for each histogram peak by computing second moments just for that peak. More peaked distributions were considered more directional.

We followed the same concept, but in order to measure orientation at multiple scales, we again used the same steerable pyramid approach, with $L_{vs}$ and $L_{hs}$ as in coarseness. From these, we define the magnitude and angle filter images:
\begin{align*}
&|L_{s}| = \sqrt{{L_{vs}}^2+{L_{hs}}^2}\\
&\angle L_{s} = tan^{-1}\bigg(\frac{L_{vs}}{L_{hs}}\bigg)
\end{align*}
We then want to ignore orientation estimates for weakly oriented regions. To find a threshold for ``orientedness'', at each scale $s$, we calculated the mean and standard deviation over all pixels of $|L_{s}|$, setting one standard deviation below the mean as our threshold $t_s$, and the removed all $\angle L_{sxy}$ such that $|L_{sxy}| < t_s$.

At each scale $s$, we binned angles (orientations) into a histogram. Rather than following the peak-counting analysis of Tamura et al.~\shortcite{tamura1978textural}, we normalized the area under the curve of the orientation distribution to 1, and calculated the entropy $e_s$ of the distribution at each scale, with the logic that an image with strong directional signal at a given scale would result in a peaky, low-entropy distribution, whereas lack of directionality would result in a uniform, high-entropy distribution. We take the minimum value $e_{min}$ for entropy over all pyramid levels, scaling such that the final range is [0,1]. Directionality is then:

\begin{align*}
Dir = 1 - 10^{\frac{e_{min}}{norm}}
\end{align*}

\subsubsection{Line-likeness} is a measure of how often in the texture line edges are present, as opposed to non-directional blobs. Tamura et al.~\shortcite{tamura1978textural} measured line-likeness as opposed to ``blob-likeness'' by measuring the co-occurrences of contrast along an edge's direction at a small range of distances. This definition of line-likeness, has a very strong correlation with directionality, such that its utility in the context of a separate global directionality measure was limited. We reasoned that, given the intent of line-likeness is to detect texture constituents such as lines that would, at larger scale, be considered directional, it was more useful to differentiate textures by a related metric, local directionality. To calculate this, we applied the directionality equation $Dir$ on tiled sub-image windows, at scales of 32x32, 64x64, and 128x128, with 100 randomly placed windows at each scale. For some textures that lack fine spatial detail, computation of directionality for small windows can give noisy results; we therefore discarded windows with a luminance variance less than 1e-5. We compute line-likeness (local directionality) by taking the mean directionality over the 100 windows for each window size (scale), followed by taking the maximum over all scales.
\begin{align*}
&Lin_{w} = \frac{1}{100} \sum_{i=1}^{100} Dir_{wi}\\
&Lin = \max_w Lin_w
\end{align*}
We chose the maximum over scales rather than the mean, with the idea that a texture that is line-like if it is line-like at any one scale.  We again exponentiated this value with base 10, and divide by a normalizing factor, giving a [0,1] range for line-likeness over all textures.

\subsubsection{Roughness} Tamura et al.~\shortcite{tamura1978textural} intended their measure of roughness to correspond to the tactile sense of roughness. They measured the sum of coarseness and contrast, but found it was poorly matched to human perception of roughness. Given roughness is typically composed of many sharp edges in a texture, we reasoned that roughness may be able to be inferred by measuring the amount of sharp edges in an image. We therefore measured roughness using the definition of 'phase congruency' \cite{kovesi2000phase}, a measure of local image energy due to phase alignment, which also detects corners and edges in images \cite{kovesi2003phase}. We used the same steerable pyramid approach as in the other descriptors, but utilizing a complex pyramid \cite{portilla2000parametric} with 6 orientations, for which the real and imaginary parts of each filter are related by the Hilbert transform. 
\begin{align*}
&L_{ros} = \Re(I \circledast \mathcal{F}_{so})\\
&L_{ios} = \Im(I \circledast \mathcal{F}_{so})
\end{align*}
For each orientation $o$ and scale $s$, we calculated the L2 norm (magnitude) of the real and imaginary pyramid images:
\begin{align*}
&L_{os} = \sqrt{{L_{ros}}^2 + {L_{ios}}^2}
\end{align*}
Our Roughness value $Rgh$ for the image, is simply the mean of this value over all orientations, scales, and pixels. Rather than exponentiating, given that roughness values tended to be condensed towards the top of the distribution, we took the log base 10 of roughness, then averaged this local energy over all pixels in each filter image and all orientations and scales. Finally, we subtracted the minimum and divided by the maximum over all textures to normalize the distribution to the range [0,1].

\begin{figure}[bt]
\centering
\includegraphics[width=1\columnwidth]{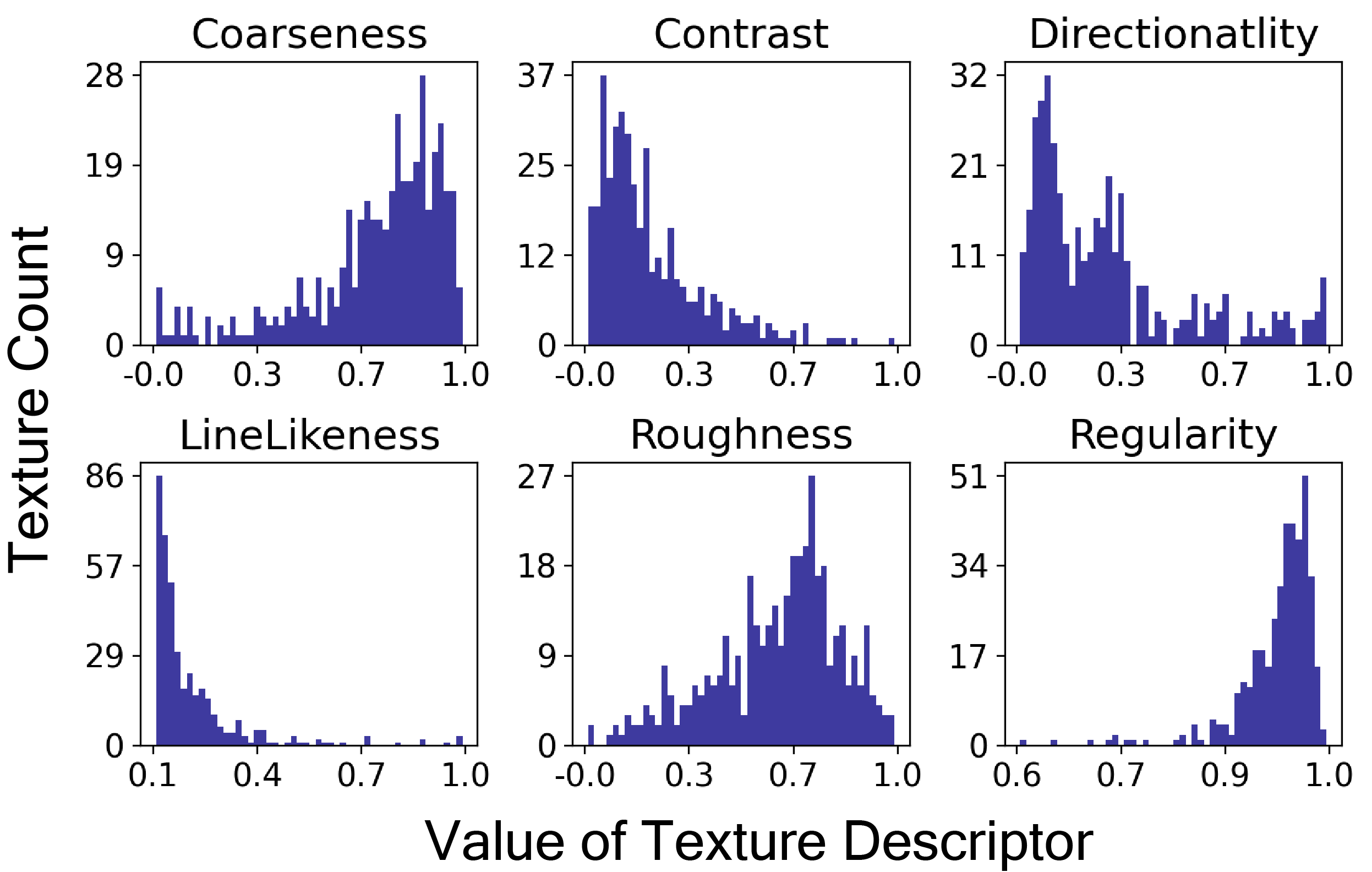}
\caption{Distribution of normalized values for each of the 6 statistics over the texture set. We applied log and exponential scaling to improve the spread for some features, and normalized such that the values for all texture descriptors in our dataset fell within the range [0,1]. Each histogram is scaled separately for optimal visibility.}
\label{fig:stat_hists}
\end{figure}

\subsubsection{Regularity} refers to patterns or statistics that repeat and/or remain constant across the image. It is somewhat related to Rao and Lohse's~\shortcite{rao1996towards} ``repetitiveness'' or Liu and Picard's~\shortcite{liu1996periodicity} ``periodicity''. We follow Tamura et al.~\shortcite{tamura1978textural} and operationalize this by measuring the variance in the other measures across space, but we generalize this calculation to multiple scales. In addition, given our improved roughness measurement, we include roughness variance in the calculation. And because we calculated regularity on subimages of varying scales, and our definition of line-likeness was directionality of subimages, we did not include line-likeness explicitly in in our regularity equations and instead relied on directionality to carry this signal. We calculated these values for 100 randomly-placed sub-image windows, repeating this random sampling at multiple window sizes $w$ (32x32, 64x64, and 128x128), and calculating each statistic's respective standard deviation $\sigma$ over the 100 windows at a given scale. Then, we took the minimum standard deviation for each given statistic over all scales. We used the minimum rather than the mean, as a texture may be highly regular at one scale, but not at others.
\begin{align*}
&\sigma Cor_{min} = min_w \sigma(Cor_w)\\
&\sigma Con_{min} = min_w \sigma(Con_w)\\
&\sigma Dir_{min} = min_w \sigma(Dir_w)\\
&\sigma Rgh_{min} = min_w \sigma(Rgh_w)
\end{align*}
Finally, we calculated Regularity as 1 minus the sum of the minimum standard deviations for each statistic over all scales. 
\begin{align*}
&Reg = 1 - \bigg( \sigma Cor_{min} + \sigma Con_{min} + \sigma Rgh_{min} + \sigma Dir_{min} \bigg)
\end{align*}

%[Justifications. E.G. much of the previous work was about visual textures, but was at least partially aimed at capturing characteristics of tactile textures.] Given that our textures were generated from PBR files, and as such we had both a top- and side- lit render for each, we also included a 'depth variability' characteristic, calculated by the ratio of contrast between the side and top lit renders: $DV = \frac{Con_{side}}{Con_{top}}$. [This would seem to raise the question of why we aren't including roughness and depth variability in our regularity measure.]
%Related to but distinct from coarseness, or overall scale of a texture, is the relative amount of high and low scale content within a texture. It is known that the power spectra of natural images follow the 'power law', meaning that their power spectrum falls approximately as $P = \frac{a}{f^{\alpha}}$. When averaged over many images, alpha is approximately 2.2 \cite{field1987relations}, but the value of this exponent varies significantly across individual images \cite{torralba2003statistics}. We measured the steepness of this falloff by for our textures, running a least-squares fit for $\alpha$ for each image. To counter over-fitting to the large magnitude power values for low frequencies, we transformed the data to log-scale before fitting.

Figure \ref{fig:texture_features} shows the least, median, and highest scoring textures for each statistical metric, according to each of these measures. We find that these features match well to perceptual expectations of texture qualities, and many are independent of material.

\begin{figure}[bt]
\centering
\includegraphics[width=1\columnwidth]{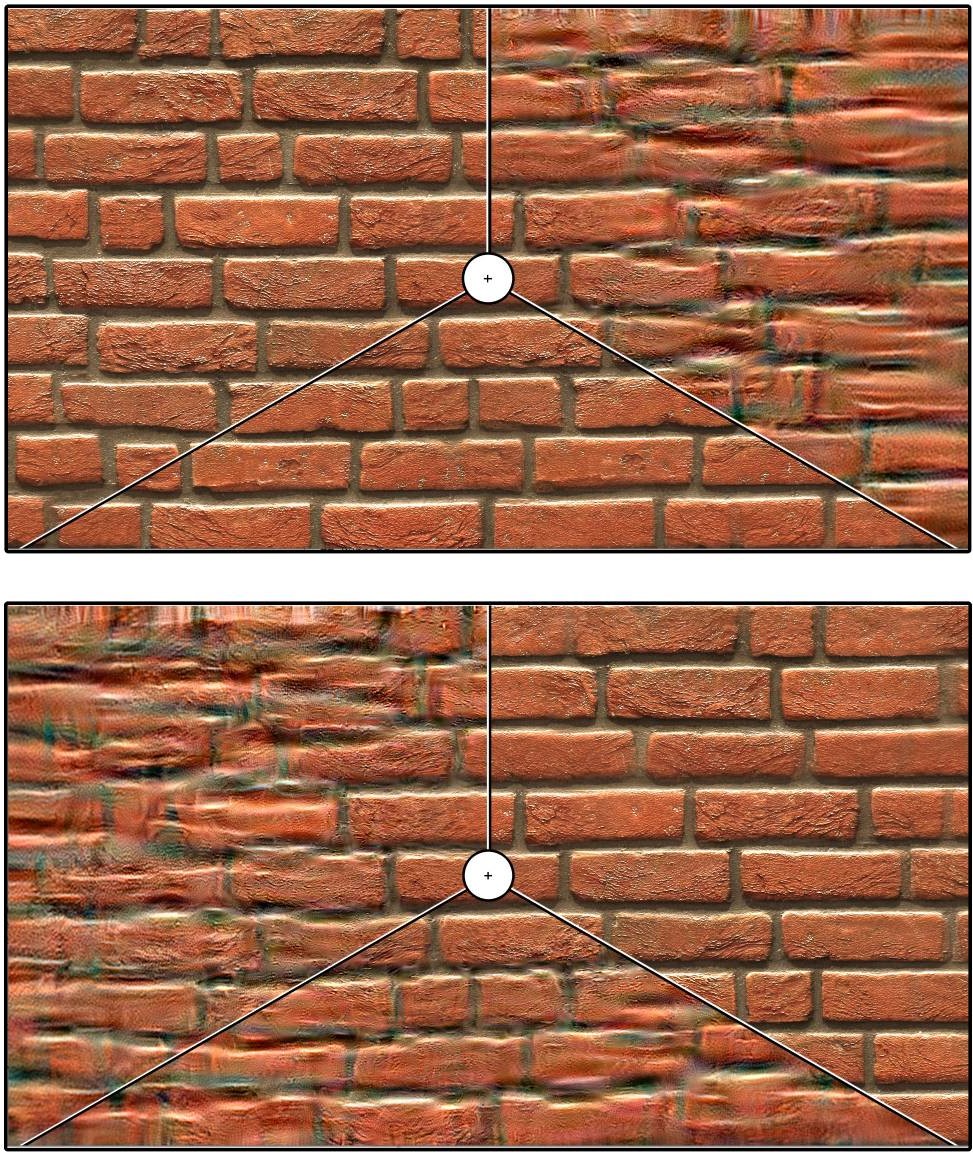}
\caption{Two example stimuli from the experiment. Each of these images would have been shown at full screen and participants were asked to fixate on the `+' in the center. For both examples shown here, the correct answer is `LEFT' indicating the lower region matches the texture on the left.}
\label{fig:stim_example}
\end{figure}

\begin{figure}
\centering
\includegraphics[width=1\columnwidth]{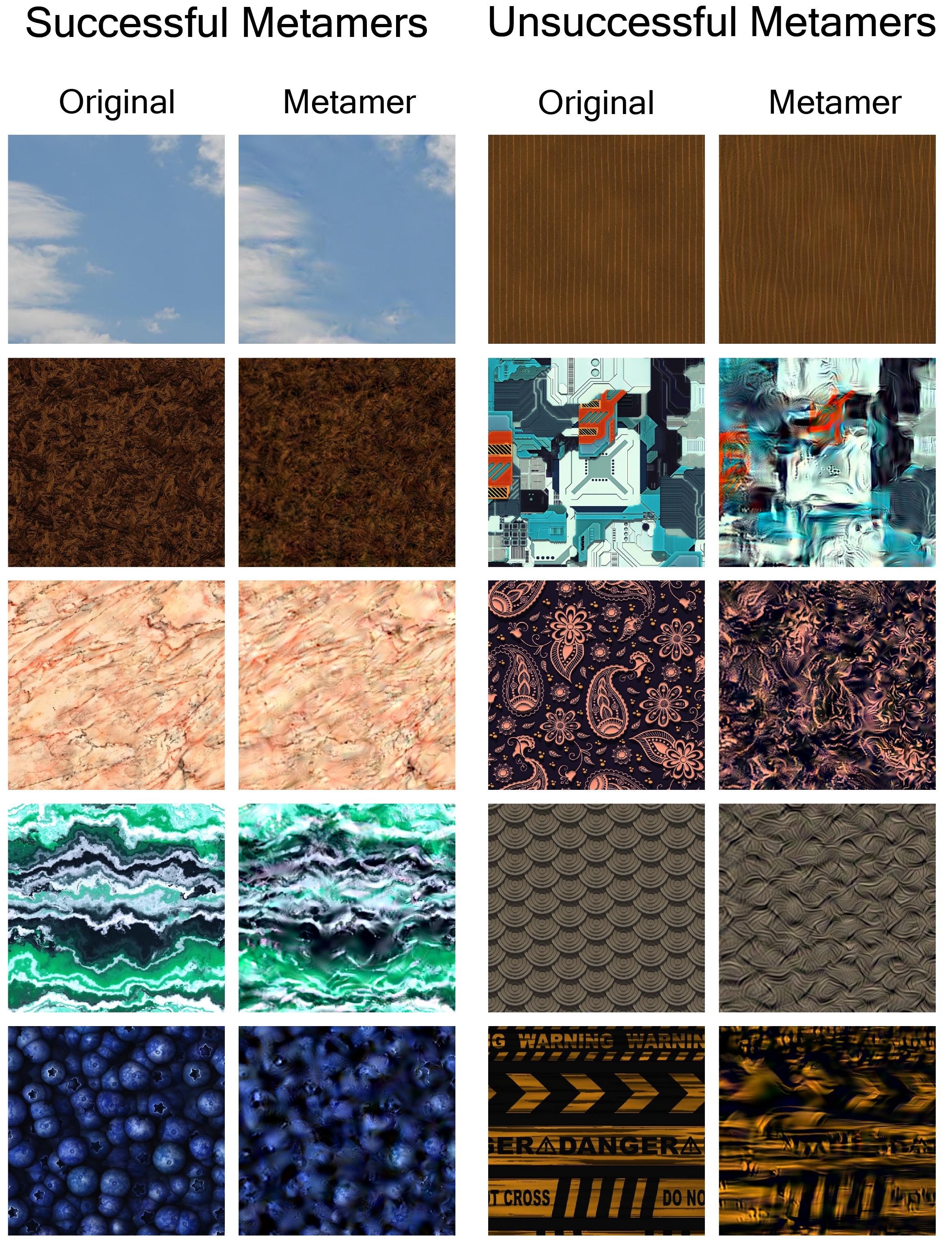}
\caption{(left) Selected examples of original and metamerized textures which succeeded, with participants at chance for detection, (right) and examples of textures that failed, with participants detecting the metamer above chance. Some of the successful metamers, such as the clouds and pink granite, appear as near-metamers even foveally, while others such as the blueberries are metamers only in the periphery. Shown are 512x512 subimages taken at half maximum eccentricity. Best viewed electronically at full-resolution.}
\label{fig:BestWorstMetamers}
\end{figure}

\section{Psychophysical Evaluation of Model}
\label{sec:evaluation}

Due to restrictions on in-person data collection, we conducted our experiment online using the Psychopy \cite{peirce2019psychopy2} and PsychoJS libraries on the Pavlovia server platform. Thirty-two adults participated in the experiment, although three were excluded based on an excessive number of timeouts and late responses (details below). The remaining twenty-nine participants ranged in age from 24 to 54 (mean of 35), and 9 were wearing glasses. Participants were asked to use a desktop monitor as a display (not a laptop screen), in a dimly lit room to maximize display visibility and contrast. Recruitment of subjects and administration of the study was conducted in accordance with the Declaration of Helsinki Ethical Principles for Medical Research Involving Human Subjects.

On each trial participants were shown a full-screen image of a texture at 1920x1080 resolution, with the screen divided into three sections and a fixation cross at the center, as depicted in Figure~\ref{fig:stim_example}. Either the left or right side of the screen showed a metamer of the texture, while the other side showed the original image. The order of the sides was randomized across textures, such that half of all trials showed the metamer on the left and vice versa. The bottom center part of the image was also independently randomized to show a metamer for half of all trials and the original for the other half. Subjects were instructed to maintain fixation at center and respond as fast as possible whether the bottom center region matched the left or right side of the screen. The matching regions could be both original texture images, or both metamers. Participants used the left and right arrow keys to indicate their response. We recorded both accuracy and reaction time. 

Participants were instructed on the experimental procedure at the start of the experiment, including a demonstration of the feedback sounds and eight practice trials. All practice trials had to be answered correctly before moving on to the main experiment. 

Trials then proceeded one after the other in blocks of 10, with rest breaks after each block to remind participants of the instructions and encourage better fixation. To encourage speedy responses, if participants failed to respond within 2.5 seconds they would hear a feedback sound indicating they had timed out, and the trial would be aborted and repeated at a random point later in the experiment. In a small percentage of the trials where participants timed out, they could have responded late and inadvertently answered the next trial incorrectly. We therefore discarded any responses within the first 200~ms of each trial.

Remote psychophysical experiments are inherently less controlled than laboratory experiments, and this may have introduced some additional noise in the data. To ensure that all participants viewed the stimuli at approximately the same visual angle, at the beginning of the experiment participants provided the physical measurements of their display and then were instructed on the necessary viewing distance to achieve 29~degrees visual angle horizontally from the center to the edge of the screen (approx. 0.9~*~screen width). This viewing distance is generally much closer than typical use, and without a chin rest subjects may have relaxed their posture away from the display. Because our model assumes a specific viewing distance and corresponding pooling region size as a function of eccentricity, such a viewing distance error might cause more of the peripheral metamer image to fall within the foveal field of view and thereby improve performance on the task. Additionally, variations in lighting environment and monitor calibration could decrease visibility, resulting in worse performance on some textures. In our case, three textures were reported to be too dark to seen by several participants and were subsequently excluded from the analysis.

Without an eye tracker we were unable to validate participants’ fixation accuracy. The stimulus duration would have allowed for several saccades, which could potentially be reflected as a combination of higher accuracy but longer reaction times. An earlier version of the experiment included a fixation task to ensure subjects were looking at the center of the screen for the entire duration of the trial, but we found that the added cognitive load made the peripheral task much too difficult. Instead we tried to optimize the experiment design to discourage eye movements by continuously maintaining the fixation marker on the screen across a rapid succession of trials. Furthermore, because accurate performance on the task required participants to allocate relatively equal attention to all three areas of the screen, we hoped that participants would be incentivized to keep their gaze at center. As a sanity check, we used an eye tracker to observe the quality of fixations for one na\"ive subject inexperienced with psychophysics experiments and found that the overall level of fixation was sufficient. 

In general, tasks requiring participants to maintain fixation while allocating attention to the periphery are very challenging, and most people improve with practice. This practice effect is somewhat but not entirely alleviated by the random order of textures across participants. On the other hand, the random texture ordering and the rapid succession of trials (intended to assist fixation) may cause very bright and very dark textures to appear in sequence, which could have lengthened reaction time due to light adaption effects.

\subsection{Analysis of Results}

Across participants, we observed a wide range of skill levels in identifying metamers, from 53\% to 79\% with an average of 64\%. We also saw a range of average reaction times across subjects, from 0.63 sec to 1.57~sec, with a mean of 1.08~sec. In general, reaction time showed a moderate correlation with percent correct ($\rho$=-0.416, p<<0) where longer reaction times correlated with lower percentage correct, as expected. As mentioned above, three participants were excluded for having too many timeouts (greater than two standard deviations from the mean). For the remaining participants the average number of timeouts was 5.5, and the average number of accidental late responses was 2.3 (defined as a reaction time less than 200~ms).

Consistent with our expectations, the distribution of mean accuracy per texture also covered a wide range, from 31\% to 100\% correct, with an average of 64\% correct. Figure~\ref{fig:BestWorstMetamers} shows examples of textures that were difficult to distinguish from their metamers (accuracy indistinguishable from chance level of 50\%) and also textures that were easy to distinguish (average accuracy above 90\%).

We first asked whether the material categories alone could predict which textures were easier than others to distinguish from their metamers. Essentially, we wanted to determine whether the accuracy distribution within each category was significantly different from a random selection across categories. We therefore calculated the average distribution of accuracy across 100 sets of 20 randomly selected textures and compared this random average to the distribution of accuracy per category using a Mann-Whitney rank test~\cite{mann1947test}. None of the distributions in any of the material categories were significantly different from random, likely due to the deliberately large variation of appearance within each category. However, on average the most difficult materials to detect were snow/smoke and painted textures, with average percent correct of 49\% and 53\% respectively, and the easiest materials to detect were plastic (81\%) and tiled textures (87\%).

Additionally, we examined the correlation between accuracy and lighting. Our expectation was that side-lit textures would on-average have higher contrast and different edge and spatial frequency distributions due to self-shadowing, and we wanted to ensure that our texture representation could be robust to these changes. We found a moderately high correlation across lighting conditions ($\rho$=0.623, p<<0); the average percent correct for each lighting condition was 64.7\% (front-lit) and 64.6\% (side-lit). A Mann-Whitney rank test showed no significant difference between accuracy distributions across lighting conditions. These results indicate that there was no systematic advantage across all textures for either lighting condition, although some metamers may have been easier to spot under one lighting condition versus the other.

We next looked at whether characteristics of the textures predict metamer success, and so turned to the seven texture descriptors described in Section~\ref{sec:descriptors}. To account for both the correlated variation across repeated measures within participants and the missing data from accidental late responses, we performed a linear mixed-effects analysis~\cite{lindstrom1988newton} of the relationship between accuracy and descriptor. Fixed effects included each of the six descriptors and all pairwise interaction terms, with a random effect of subject. None of the descriptors showed a significant main effect, but we found two pairwise interactions significant at the p<0.01 level: Contrast \& Roughness ($\beta$=-0.673, SE=0.214, z(11375)=-3.142, p=0.002), and Coarseness \& Roughness ($\beta$=0.589, SE=0.135, z(11375)=4.355, p<<0.000).

\begin{figure}
\centering
\includegraphics[width=1.0\columnwidth]{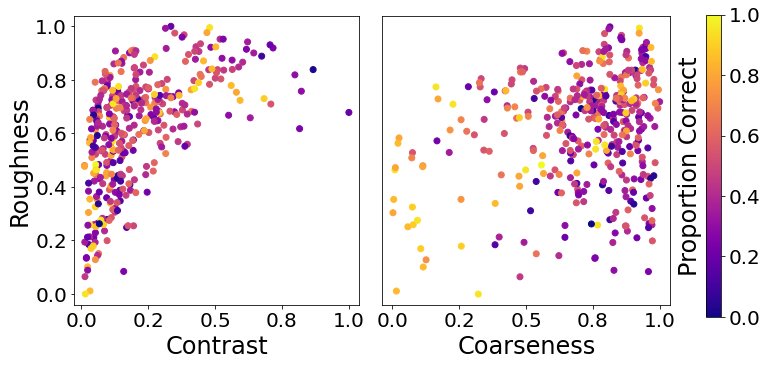}
\caption{Relationship between accuracy (proportion correct) and texture descriptors for the two interactions we observed: Contrast vs Roughness and Coarseness vs Roughness. Darker colors indicate a better metamer match, while lighter colors indicate the metamer was easily detectable. A high positive correlation on the x-axis with no interaction, for example, would show a fade from dark to light along the x-axis but a random distribution of colors along the y-axis. However, a strong interaction would show a cluster of dark (or light) colors only in one corner of the plot (e.g. top-right for Contrast and bottom-left for Coarseness).}
\label{fig:statistics}
\end{figure}

Visual inspection of these two relationships, shown in Figure~\ref{fig:statistics}, illustrate that the best metamer matches were achieved by textures with high contrast and high roughness, while the worst metamer matches correlated with low coarseness and low roughness. It is interesting in the contrast/roughness interaction in particular, that high contrast (in the context of high roughness) is associated with low-detectibility (successful metamer), given that high contrast in a visual stimulus is typically associated with higher visibility; one might expect that any artifacts in the metamer process would be more detectable. Perhaps high contrast images, in the context of roughness, provide a strong overall structure, which, if maintained, serves to mask smaller scale and low contrast variations that would otherwise be giveaways. This interpretation is consistent with low roughness and low coarseness to be predictors of unsuccessfull metamers, as these textures, although they may not necessarily have low contrast, may not have this strong overall structure capable of such masking.

\begin{figure}
\centering
\includegraphics[width=1.0\columnwidth]{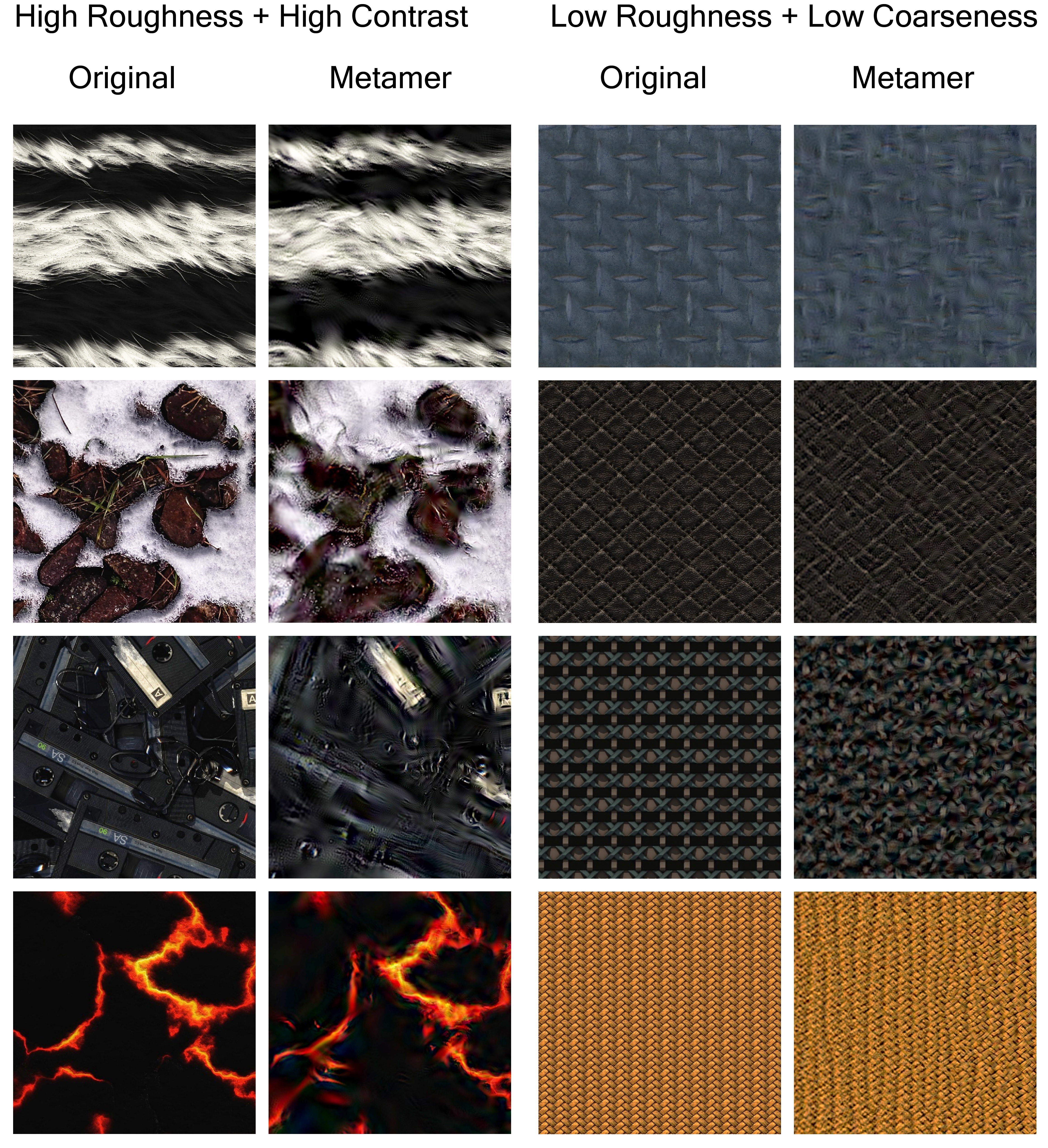}
\caption{Textures corresponding to interactions between the texture properties identified in Figure \ref{fig:statistics}. High roughness combined with high contrast (left) typically results in successful metamers which avoid detection. Low roughness combined with low coarseness (right) typically result in unsuccessful metamers which are detectable. Contrast enhanced for better viewing at a small scale.  Best viewed electronically at full-resolution.} 
\label{fig:TexureLowHighProps}
\end{figure}

In addition, Figure~\ref{fig:TexureLowHighProps} shows a few examples of textures at the extreme ends of these distributions. Based on visual inspection, one notable difference between the hard to detect metamers on the left and the easy to detect metamers on the right is the very high degree of \emph{both} local and global regularity in the textures on the right. Highly regular textures may lead to poor metamers because if one detects even a small deviation from regularity that provides a cue to distinguish that image from the original. Why, then, did our measure of regularity not distinguish been good and bad metamers? Our definition of regularity merely required textures to be regular at \emph{any one} scale; any irregularity at another scale might reduce the cues distinguishing the original from metamer image, leading to a more successful metamer. Regardless, textures with cross-scale regularity do not appear to make good metamers using our synthesis procedure. It is possible that that our model fails to account for some longer-range structure that the visual system is able to perceive, or this may simply be due to a failure of the synthesis procedure to fully converge in the case of textures with very low randomness. Fortunately, the flexibility of our model allows us to test many variations of statistics with ease, and we are continuing to explore better ways to synthesize metamers even for these particularly difficult textures.

\section{Conclusions and Future Work}

We have reviewed the summary statistic encoding model of peripheral vision advocated by many vision scientists. Their work is in progress and does not yet provide a ``drop in'' model that graphics practitioners can apply in, for example, a peripherally aware renderer or 360 video encoder.  In particular, the spatial summary statistics are not yet final, and there is not yet a temporal model.  Thus our paper is mainly a call to action, because we believe that the summary statistics approach is the most promising one for delivering a loss function that would be needed for an efficient peripheral renderer or encoder.  One of the main practical barriers to vision scientists making progress in that summary statistics work is that previous computational approaches have led to very long run times and relatively small images.  We have described a different computational approach, based on the type of data-flow computation used in most deep learning tool-kits, that greatly improves the boundaries of what data can be run, and how quickly it can be processed (the code in PyTorch for our approach will be distributed with the paper).  Further, we have done a psychophysical study to characterize some of the texture domains where our version of the summary statistics model applies.

Our work can likely be applied immediately in some domains such as texture synthesis, but the biggest payoffs, in animated systems, will need time dependence to be accounted for.  This may or may not prove to be well-modeled by temporal or spatio-temporal statistics, though early evidence suggests that crowding occurs across time as well as space, and that temporal crowding has similar characteristics to spatial crowding \cite{bex2005spatial, holcombe2009seeing}.  We think the tools we have introduced in this paper will enable vision science researchers to explore that space of possibilities; the previous tools came with practical barriers to such work both in terms of difficulty testing new statistics and computational intensity.  Another avenue that may prove fruitful is for graphics researchers to use the spatial summary statistics from this paper in conjunction with empirical work on motion; such an approach has proven very fruitful in non-peripheral contexts using temporal anti-aliasing.  In general, we believe this will be a very productive area of collaboration between graphics researchers and vision scientists, as many other areas have in the past.

\begin{acks}
Acks (invisible for review)
\end{acks}

\bibliographystyle{ACM-Reference-Format}
\bibliography{main}
\end{document}